\documentclass[journal]{IEEEtran}
\usepackage{graphicx}
\usepackage{subfigure}
\usepackage{subeqnarray}
\usepackage{bm}
\usepackage{amsfonts}
\usepackage{amsthm}
\usepackage{amssymb}
\usepackage{makecell}
\usepackage{calc}
\usepackage{color}
\usepackage{mathrsfs}
\usepackage{flushend}
\usepackage{url}
\usepackage{caption}
\usepackage{subfloat}
\usepackage{color}
\usepackage{textcomp}

\usepackage[outdir=./]{epstopdf}

\usepackage{cite}

\usepackage{booktabs}

\newtheorem{remark}{Remark}

\ifCLASSINFOpdf
\else
\fi
\usepackage{amsmath}
\usepackage{amsthm}

\usepackage{algorithm}
\usepackage{algorithmic}
\ifCLASSOPTIONcompsoc
 \usepackage[caption=false,font=normalsize,labelfont=sf,textfont=sf]{subfig}
\else
 \usepackage[caption=false,font=footnotesize]{subfig}
\fi
\begin{document}
\captionsetup[figure]{labelformat={default},labelsep=period,name={Fig.}}
\title{Secure Dual-Functional Radar-Communication Transmission: Exploiting Interference for Resilience Against Target Eavesdropping}
%
%
%

\author{Nanchi Su,~\IEEEmembership{Student~Member,~IEEE}, Fan Liu,~\IEEEmembership{Member,~IEEE}, Zhongxiang Wei,~\IEEEmembership{Member,~IEEE}, Ya-Feng Liu,~\IEEEmembership{Senior~Member,~IEEE}, and Christos Masouros,~\IEEEmembership{Senior~Member,~IEEE}

\thanks{{This work was supported by the Engineering and Physical Sciences Research Council projects EP/R007934/1 and EP/S026622/1, and the China Scholarship Council (CSC). \textit{(Corresponding author: Fan Liu.)}}}
\thanks{N. Su and C. Masouros are with the Department of Electronic and Electrical Engineering, University College London, London WC1E 7JE, U.K. (e-mail: nanchi.su.18@ucl.ac.uk, chris.masouros@ieee.org).}
\thanks{F. Liu is with the Department of Electrical and Electronic Engineering, Southern University of Science and Technology, Shenzhen 518055, China (e-mail: liuf6@sustech.edu.cn).}
\thanks{Z. Wei is with the College of Electronic and Information Engineering, Tongji University, Shanghai, China. (email: z\_wei@tongji.edu.cn)}
\thanks{Y.-F. Liu is with the State Key Laboratory of Scientific and Engineering Computing, Institute of Computational Mathematicsand Scientific/Engineering Computing, Academy of Mathematics and Systems Science, Chinese Academy of Sciences, Beijing 100190, China (e-mail: yafliu@lsec.cc.ac.cn).}
}

\maketitle

\begin{abstract}
   We study security solutions for dual-functional radar communication (DFRC) systems, which detect the radar target and communicate with downlink cellular users in millimeter-wave (mmWave) wireless networks simultaneously. Uniquely for such scenarios, the radar target is regarded as a potential eavesdropper which might surveil the information sent from the base station (BS) to communication users (CUs), that is carried by the radar probing signal. Transmit waveform and receive beamforming are jointly designed to maximize the signal-to-interference-plus-noise ratio (SINR) of the radar under the security and power budget constraints. We apply a Directional Modulation (DM) approach to exploit constructive interference (CI), where the known multiuser interference (MUI) can be exploited as a source of useful signal. Moreover, to further deteriorate the eavesdropping signal at the radar target, we utilize destructive interference (DI) by pushing the received symbols at the target towards the destructive region of the signal constellation. Our numerical results verify the effectiveness of the proposed design showing a secure transmission with enhanced performance against benchmark DFRC techniques.
\end{abstract}


%
\IEEEpeerreviewmaketitle

\section{Introduction}
%
%
%
%
\subsection{Background and motivation}
\IEEEPARstart{W}{ireless} spectrum is getting increasingly congested due to the tremendous growth of wireless connections and mobile devices, which results in high auction price of the available frequency bands. According to \cite{Survey2018}, the Spanish government raised a total of \texteuro 438 million for the sale of 5G frequencies. On the other hand, the government of South Korea paid \$3.3 billion for the 3.5 GHz and 28 GHz bands in 5G network. To address the increasing need for extra spectrum, the radar bands, which are largely overlapped with those of major communication applications, have been envisioned as potentially exploitable spectral resources. In fact, given the overlapped frequencies, as well as the more and more similar RF front-end designs between radar and communications, the shared use of the spectrum or even the hardware platform between both functionalities becomes a promising solution to improve the efficiency and reduce the costs. This has given rise to the development of the Dual-functional Radar-Communication (DFRC) system in recent years \cite{zhang2017waveform, liu2020joint, liu2018mu, liu2018toward, yuan2020spatio}. In many emerging applications, DFRC systems are expected to meet the demand for location-awareness as a new paradigm, for example, in intelligent transportation systems \cite{hassanien2019dual}.
\\\indent In DFRC systems, the transmitted waveform is specifically designed as to serve for both purposes of target sensing and wireless communication, which raises unique security challenges. Intuitively, the radar beampattern is designed to concentrate the radiation power towards the direction of targets of interest so as to improve the detection performance. Since the probing DFRC signal also carries information for the communication users, the target, as a potential eavesdropper, e.g., an unauthorized vehicle or UAV, could readily surveil the information intended for communication users (CUs). To this end, new physical layer (PHY) security solutions are required for the dual functional operation in security-critical DFRC designs.
\\\indent Methods to secure the wireless communication systems are widely investigated over the past decades. Pioneered by Wyner \cite{wyner1975wire}, beamformer and precoder are designed to ensure the quality-of-service (QoS) at legitimate users while limiting the signal strength received at the potential eavesdroppers \cite{lv2015secrecy, gong2016millimeter, liu2014secrecy, yang2014algorithms}, which aims to maximize the signal-to-interference-plus-noise ratio (SINR) difference between the two types of users, and accordingly yields an optimal secrecy rate (SR). On top of that, artificial noise (AN) is generated to further deteriorate the received signals at eavesdroppers \cite{zhang2018artificial, kong2019robust, wang2017artificial, su2020secure, 8903316}. AN-aided scheme is proved to be efficient especially when the channel statement information (CSI) of eavesdroppers is unknown or partially known to the base station (BS) \cite{wang2020intelligent, he2017joint}.
\\\indent In conventional beamforming designs, AN indeed degrades SINR at both CUs and eavesdroppers, which requires higher power budget to ensure the QoS. In view of the redundant power consumption caused by AN, directional modulation (DM) has attracted growing research attentions as an emerging hardware efficient approach to secure wireless communication systems in recent years \cite{wei2020secure, yan2015optimization,baghdady1990directional}. The DM transmitter sends confidential information to the CUs such that the malicious eavesdroppers cannot intercept the transmitted messages \cite{shu2019directional}. Unlike the SR based methods, DM technique adjusts the amplitude and phase of the symbols at the users of interest directly while scrambling the symbols in other undesired directions, which implies that the modulation happens at the antenna level instead of at the baseband level. As a result, a low symbol error rate (SER) can be endorsed at the CUs, while the received symbols of the eavesdropper are randomized in the signal constellation. Since the expensive and power-consuming radio frequency (RF) chains and digital-to-analog converter (DAC) deployed in conventional beamforming design are not required, the DM based scheme is efficient on aspects of both cost and energy. The DM approach is based on the pronciples of exploiting constructive interference (CI) \cite{masouros2014vector, masouros2015exploiting, liu2018mimo, khandaker2018secure}, where the received signal is not necessary to be aligned with the intended symbols, but is pushed away from the detection thresholds of the signal constellation.
\\\indent In this relevant line of CI research, recent studies focus on exploiting CI through symbol-level precoding, which exploits known multiuser interference (MUI) as useful power by pushing the received signal away from the detection bound of the signal constellation. Also, it is provable that CI-based precoding designs benefit the data secrecy. In particular, the CI and AN can be jointly exploited to design secure beamformer under the assumption of perfect or imperfect CSI \cite{khandaker2018secure,khandaker2018constructive}, which was proved to outperform the conventional AN-aided secrecy optimization. In addition to increasing the secrecy, the generated AN was exploited to be constructive to energy harvesting in \cite{khandaker2018secure}. AN-aided CI precoding designs were proposed in \cite{wei2019interference, wei2019robust, 8928979}, where a deterministic robust optimization algorithm was presented in \cite{wei2019interference} and a probabilistic optimization method was presented in \cite{wei2019robust}, respectively. Furthermore, the work of \cite{8928979} expanded the scenario to more practical cases where the CSI of eavesdropper is totally unknown. In \cite{wei2020secure}, practical transmitter designs were exploited when the CUs' channel is correlated with or without the eavesdropper's channel. We note that while all the above approaches are designed for the classical PHY security scenario involving legitimate users and external eavesdroppers, none of these apply to the unique DFRC scenarios where the target of interest may be a potential eavesdropper.
\\\indent To address the security issue raised in the DFRC systems, in \cite{deligiannis2018secrecy}, the MIMO radar was designed to transmit a mixture of two different signals, including desired information for the intended users and a pseudorandom distortional waveform to confuse the eavesdropper, both of which are used for detecting the target. In this context, several optimizations were designed, namely target return SINR maximization, transmit power minimization, and SR maximization, where the former two designs keep the SR above a given threshold. In \cite{chalise2018performance}, a unified system including passive radar and communication system has been studied. To ensure the SR at CUs, the optimization problem was designed to maximize the SINR at passive radar with an SR threshold constraint. Furthermore, an AN-aided method deployed in DFRC systems was proposed in \cite{su2020secure}, where the BS serves communication users and detects a target simultaneously. To secure the communication data via optimized SR, the SNR was minimized at the target while ensuring the SINR at each desired user.
\\\indent To the best of our knowledge, all the existing studies on DFRC security are based on SR maximization, with the assumption of Gausssian symbol transmission and perfect or imperfect CSI knowledge. To address DFRC security in broader scenarios, it is worth studying the CI based waveform design for the reason that a) MUI is commonly treated as a detrimental impact that needs to be mitigated, while it becomes beneficial and further contributes to the useful signal power in CI design; b) CI based precoding can support a larger number of data streams with a significantly improved SER performance \cite{li2020multiplexing}.
\subsection{Contributions}
We propose several designs, which aim at maximizing the receive SINR of radar in secure DFRC systems. Specifically, we consider a MIMO DFRC BS which serves CUs and detects a point-like target simultaneously, where the transmit waveform and the receive beamformer are jointly designed to improve PHY security following the CI approach. Note that the target is treated as a potential eavesdropper. As a further consideration on communication data secrecy, MUI is designed to be constructive at the CUs, while disrupting the data at the radar target, which deteriorates the target receive signals and thus increases the SER at the target. Throughout this paper, the proposed problems above are firstly studied in an ideal scenario where the target location is known to the BS, and are then extended to the more practical case where the location is uncertain to the BS.
\\\indent Within this scope, the contributions of our work are summarized as follows:
\begin{itemize}
    \item We design the transmit waveform and receive beamformer jointly for the secure DFRC system, where the DM technique is employed to maximize the received SINR of the radar system under the constraints of power budget and CI for security.
    \item We propose a fractional programming (FP) algorithm to solve the radar SINR maximization problem, and compare the resulting performance with benchmark techniques, and alternative solvers including semidefinite relaxation (SDR), and successive QCQP (SQ) methods.
    \item We investigate the problem under the practical condition of target location uncertainty, where the DFRC waveform is designed to maximize the minimum radar SINR within a given angular interval that the targets might fall into.
    \item We further consider an advanced secure CI design for the proposed DFRC system, where the MUI is designed to be constructive to CUs, while destructive to the target.
\end{itemize}
\subsection{Organization}
This paper is organized as follows. Section II gives the system model. The waveform optimization problem is designed with the guarantee of PHY security by adopting CI method in Section III and Section IV, when the target location is known to the BS perfectly or imperfectly, respectively. In Section V, PHY security is further considered by constructing the received signal at the target into the destructive region. Section VI provides numerical results, and Section VII concludes the paper.
\\\indent \emph{Notations}: Unless otherwise specified, matrices are denoted by bold uppercase letters (i.e., $\mathbf{X}$), vectors are represented by bold lowercase letters (i.e., $\mathbf{x}$), and scalars are denoted by normal font (i.e., $\alpha$). Subscripts indicate the location of the entry in the matrices or vectors (i.e., $s_{i,j}$ and $l_n$ are the $(i,j)$-th and the \emph{n}-th element in $\mathbf{S}$ and $\mathbf{l}$, respectively). $\operatorname{tr}\left(\cdot\right)$ and $\operatorname{vec}\left(\cdot\right)$ denote the trace and the vectorization operations. $\left(\cdot\right)^T$, $\left(\cdot\right)^H$ and $\left(\cdot\right)^*$ stand for transpose, Hermitian transpose and complex conjugate of the matrices, respectively. $\operatorname{diag}\left(\cdot\right)$ represents the vector formed by the diagonal elements of the matrices and ${\text{rank}}\left(  \cdot  \right)$ is rank operation. $\left\| \cdot\right\|$, $\left\| \cdot\right\|_{\infty}$ and $\left\| \cdot\right\|_F$ denote the $l_2$ norm, infinite norm and the Frobenius norm respectively. $\mathbb{E}\left\{ \cdot  \right\}$ denotes the statistical expectation.

\begin{figure}
    \centering
    \subfigure[]{
    \includegraphics[width=0.47\columnwidth]{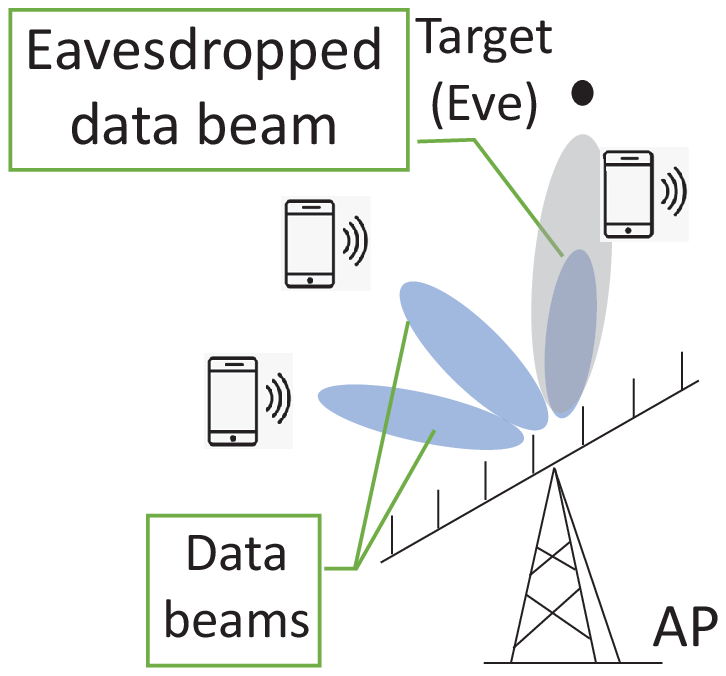}}
    \subfigure[]{
    \includegraphics[width=0.47\columnwidth]{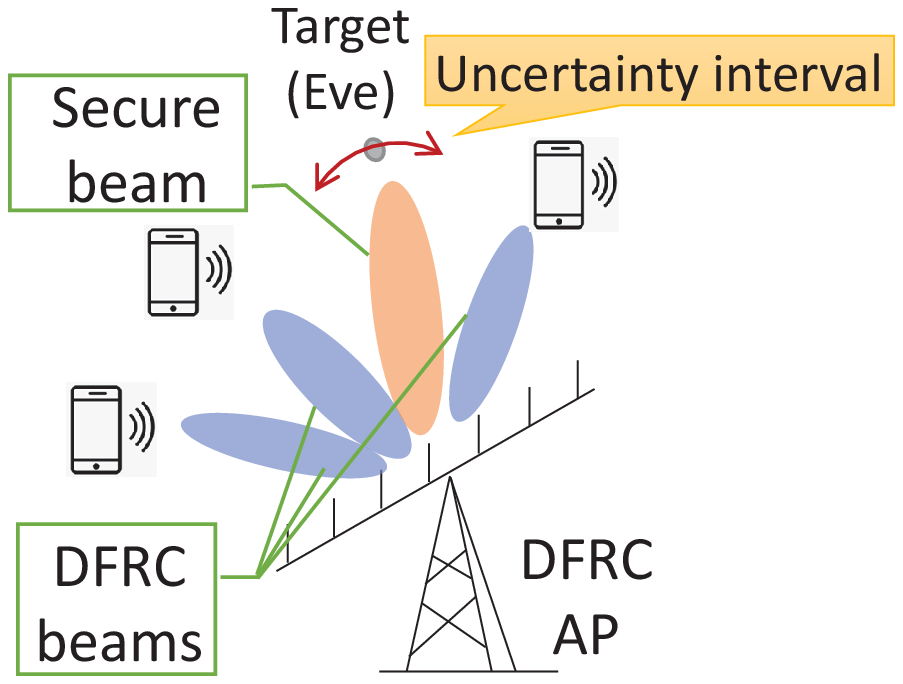}}
    \captionsetup{font={footnotesize}}
    \caption{(a) DFRC System imposed potential eavesdropper. (b) Secure DFRC system.}
    \label{fig.1}
\end{figure}
\section{System Model}
We consider a DFRC MIMO system with a BS equipped with ${N_T}$ transmit antennas and ${N_R}$ receive antennas, which is serving \emph{K} single-antenna users and detecting a point-like target simultaneously. As shown in Fig.1, the target can be regarded as a potential eavesdropper which might intercept the information sent from the BS to legitimate users. Due to the existence of $I$ clutter sources, the target return is interfered at the BS's receiver. Additionally, the communication channel is considered to be a narrowband slow time-varying block fading Rician fading channel. Based on the assumptions above, below we elaborate on the radar and communication signal models.
\subsection{Radar Signal Model}
Let ${\mathbf{x}} \in {\mathbb{C}^{{N_T} \times 1}}$ denote the transmit signal vector, the received waveform at the target is given as
\begin{equation}\label{eq1}
    {\mathbf{r}} = \underbrace {{\alpha _0}{\mathbf{U}}\left( {{\theta _0}} \right){\mathbf{x}}}_{\text{signal}} + \underbrace {\sum\limits_{i = 1}^I {{\alpha _i}{\mathbf{U}}\left( {{\theta _i}} \right)} {\mathbf{x}}}_{\text{signal - dependent{\text{ c}}lutter}} + \underbrace{\mathbf{z}}_{\text{noise}},
\end{equation}
where ${\alpha _0}$ and ${\alpha _i}$ denote the complex amplitudes of the target and the \emph{i}-th interference source, ${\theta _0}$ and ${\theta _i}$ are the angle of the target and the \emph{i}-th signal-dependent clutter source, respectively, and ${\mathbf{z}} \in {\mathbb{C}^{{N_R} \times 1}}$ is the additive white Gaussian noise (AWGN) vector, with the variance of $\mathbf{\sigma} _R^2$. ${\mathbf{U}}\left( {{\theta}} \right)$ is the steering matrix of uniform linear array (ULA) antenna with half-wavelength spaced element, defined as
\begin{equation}\label{eq2}
    {\mathbf{U}}\left( \theta  \right) = {{{\mathbf{a}}_r}\left( \theta  \right){\mathbf{a}}_t^T\left( \theta  \right)},
\end{equation}
where ${{\mathbf{a}}_t}\left( \theta  \right) = \frac{1}{{\sqrt {{N_T}} }}{\left[ {1,{e^{ - j\pi \sin \theta }}, \cdots ,{e^{ - j\pi \left( {{N_T} - 1} \right)\sin \theta }}} \right]^T}$ and ${{\mathbf{a}}_r}\left( \theta  \right) = \frac{1}{{\sqrt {{N_R}} }}{\left[ {1,{e^{ - j\pi \sin \theta }}, \cdots ,{e^{ - j\pi \left( {{N_R} - 1} \right)\sin \theta }}} \right]^T}$. Then, the output of the filter can be given as
\begin{equation}\label{eq3}
\begin{split}
    {r_f} &= {{\mathbf{w}}^H}{\mathbf{r}}\\
          &= {\alpha _0}{{\mathbf{w}}^H}{\mathbf{U}}\left( {{\theta _0}} \right){\mathbf{x}} + \sum\limits_{i = 1}^I {{\alpha _i}{{\mathbf{w}}^H}{\mathbf{U}}\left( {{\theta _i}} \right)} {\mathbf{x}} + {{\mathbf{w}}^H}{\mathbf{z}},
\end{split}
\end{equation}
where ${\mathbf{w}}\in {\mathbb{C}^{{N_R} \times 1}}$ denotes the receive beamforming vector.
Accordingly, the output SINR can be expressed as
\begin{equation}\label{eq4}
\begin{split}
    {\text{SIN}}{{\text{R}}_{rad}} &= \frac{{{{\left| {{\alpha _0}{{\mathbf{w}}^H}{\mathbf{U}}\left( {{\theta _0}} \right){\mathbf{x}}} \right|}^2}}}{{{{\mathbf{w}}^H}\sum\limits_{i = 1}^I {{{\left| {{\alpha _i}} \right|}^2}{\mathbf{U}}\left( {{\theta _i}} \right){\mathbf{x}}{{\mathbf{x}}^H}{{\mathbf{U}}^H}\left( {{\theta _i}} \right){\mathbf{w}} + {{\mathbf{w}}^H}{\mathbf{w}}\sigma _R^2} }} \hfill \\
    {\text{           }}
    &= \frac{{\mu {{\left| {{{\mathbf{w}}^H}{\mathbf{U}}\left( {{\theta _0}} \right){\mathbf{x}}} \right|}^2}}}{{{{\mathbf{w}}^H}\left( {{\mathbf{\Sigma }}\left( {\mathbf{x}} \right) + {{\mathbf{I}}_{{N_R}}}} \right){\mathbf{w}}}}, \hfill \\
\end{split}
\end{equation}
where $\mu  = {{ {{{\left| {{\alpha _0}} \right|}^2}} } \mathord{\left/
 {\vphantom {{\mathbb{E}\left[ {{{\left| {{\alpha _0}} \right|}^2}} \right]} {\sigma _R^2}}} \right.
 \kern-\nulldelimiterspace} {\sigma _R^2}}$, ${\mathbf{\Sigma }}\left( {\mathbf{x}} \right) = \sum\limits_{i = 1}^I {{b_i}{\mathbf{U}}\left( {{\theta _i}} \right){\mathbf{x}}{{\mathbf{x}}^H}{{\mathbf{U}}^H}\left( {{\theta _i}} \right)} $, and ${b_i} = {{ {{{\left| {{\alpha _i}} \right|}^2}} } \mathord{\left/
 {\vphantom {{\mathbb{E}\left[ {{{\left| {{\alpha _i}} \right|}^2}} \right]} {\sigma _R^2}}} \right.
 \kern-\nulldelimiterspace} {\sigma _R^2}}$.
\\\indent Since $\mathbf{x}$ is the intended information signal, the received signal at target (eavesdropper's receiver) can be given as
\begin{equation}\label{eq5}
    {y_R} = {\alpha _0}{\mathbf{a}}_t^H\left( {{\theta _0}} \right){\mathbf{x}} + e,
\end{equation}
where $e \sim \mathcal{C}\mathcal{N}\left( {0,\sigma _T^2} \right)$ denotes the AWGN. Then, eavesdropping SNR at radar target can be expressed as
\begin{equation}\label{eq6}
    {\text{SN}}{{\text{R}}_{T}} = \frac{{{{{\left| {{\alpha _0}{\mathbf{a}}_t^H\left( {{\theta _0}} \right){\mathbf{x}}} \right|}^2}} }}{{\sigma _T^2}}.
\end{equation}
\subsection{Communication Signal Model}
The received signal at the \emph{k}-th CU can be written as
\begin{equation}\label{eq7}
    {{y}_k} = {{\mathbf{h}}_k^H}{\mathbf{x}} + {{n}_k},
\end{equation}
where ${{\mathbf{h}}_k} \in {\mathbb{C}^{{N_T} \times 1}}$ denotes the multiple-input-single-output (MISO) channel vector between the BS and the \emph{k}-th CU. Similarly, ${{n}_k}$ is the AWGN of the CU $k$ with the variance of $\sigma _{C_k}^2$. We assume that $\mathbf{h}_k$ is a slow time-varying block Rician fading channel, i.e., the channel is constant in a block but varies slowly from one block to another. Thus, the channel vector of the $k$-th user can be expressed as a combination of a deterministic strongest line-of-sight (LoS) channel vector and a multiple-path scattered channel vector, which is expressed as
\begin{equation}\label{eq8}
    {{\mathbf{h}}_k} = \sqrt {\frac{{{v_k}}}{{1 + {v_k}}}} {\mathbf{h}}_{L,k}^{{\text{LoS}}} + \sqrt {\frac{1}{{1 + {v_k}}}} {\mathbf{h}}_{S,k}^{{\text{NLoS}}},
\end{equation}
where $v_k>0$ is the Rician $K$-factor of the $k$-th user, ${\mathbf{h}}_{L,k}^{{\text{LoS}}} = \sqrt {{{N_T}}} {\mathbf{a}_t}\left( {{\omega _{k,0}}} \right)$ is the LoS deterministic component. ${\mathbf{a}}\left( {{\omega _{k,0}}}\right)$ denotes the array steering vector, where ${\omega _{k,0}} \in \left[ {{\text{ - }}\frac{\pi }{2}, \frac{\pi }{2}} \right]$ is the angle of departure (AOD) of the LoS component from the BS to user $k$ \cite{hu2019cluster, zhao2017tone}. The scattering component ${\mathbf{h}}_{S,k}^{{\text{NLoS}}}$ can be expressed as ${\mathbf{h}}_{S,k}^{{\text{NLoS}}} = \sqrt {\frac{{{N_T}}}{L}} \sum\limits_{l = 1}^L {{c_{k,l}}{\mathbf{a}_t}\left( {{\omega _{k,l}}} \right)} $, where $L$ denotes the number of propagation paths, ${c_{k,l}} \sim \mathcal{C}\mathcal{N}\left( {0,1} \right)$ is the complex path gain and ${\omega _{k,l}} \in \left[ {{\text{ - }}\frac{\pi }{2}, \frac{\pi }{2}} \right]$ is the AOD associated to the $\left( k,l \right)$-th propagation path.
\\\indent Additionally, we note that the intended symbol varies at a symbol-by-symbol basis in CI precoding designs. Let $s_k$ denote the intended symbol of the $k$-th CU, which is $M$-PSK modulated. To this end, we define ${s_k} \in {\mathcal{A}_M}$, where ${\mathcal{A}_M} = \left\{ {{a_m} = {e^{j\left( {2m - 1} \right)\phi }},m = 1, \cdots ,M} \right\}$, $\phi  = {\pi  \mathord{\left/{\vphantom {\pi  M}} \right.\kern-\nulldelimiterspace} M}$, and $M$ denotes the modulation order.
\begin{figure}
    \centering
    \subfigure[]{
    \includegraphics[width=0.45\columnwidth]{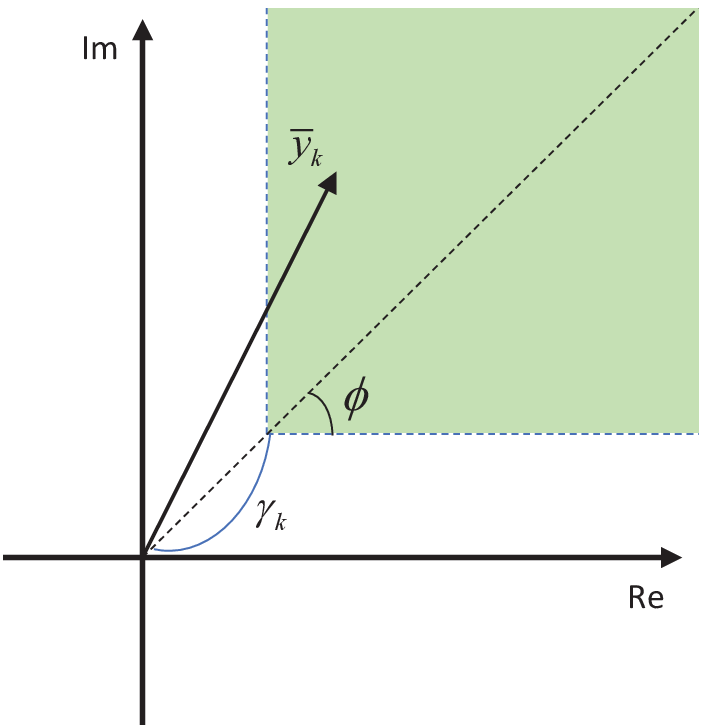}}
    \subfigure[]{
    \includegraphics[width=0.45\columnwidth]{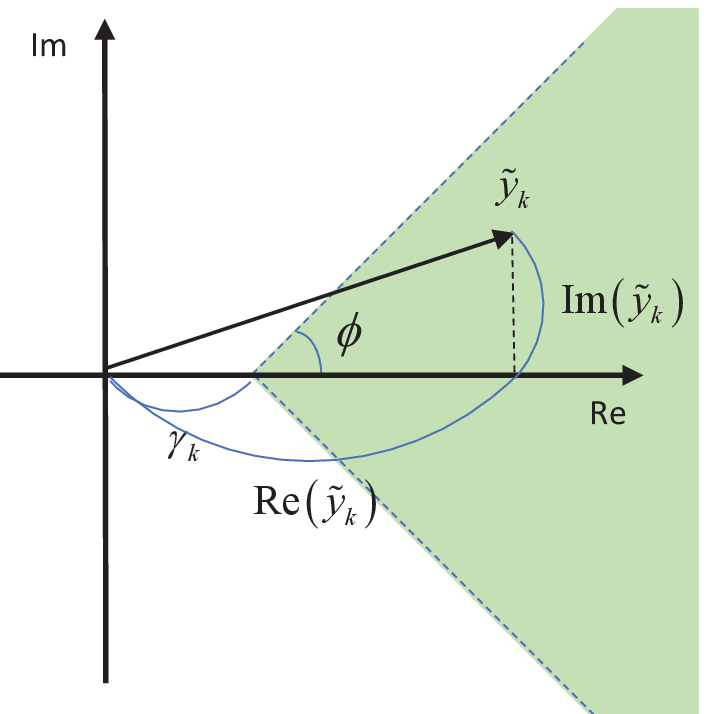}}
    \captionsetup{font={footnotesize}}
    \caption{QPSK illustration. (a) Relaxed phase DM. (b) Rotation by $\arg \left( {s_k^*} \right)$.}
    \label{fig.2}
\end{figure}
\section{${\text{SIN}}{{\text{R}}_{rad}}$ Maximization With Known Target Location}
With the knowledge of precise target location, in this section, we design the transmit waveform aiming at maximizing the received radar SINR and subject to the information secrecy constraint in the wireless communication system deploying CI method. For clarity, we remark here that the known target location is quite a typical assumption in the radar literature, especially for target tracking algorithm designs. This can be interpreted as to optimize the transmit waveform and receive beamformer towards a specific direction of interest, or to track the movement of the target with predicted location information inferred from the previous estimates. Note that this also applies to the clutter sources, whose angles are assumed to be pre-estimated.
\\\indent In light of the above system setting, we then propose two algorithms to tackle the optimization problem, namely, the successive QCQP (SQ) method proposed in Section III-B and the FP method proposed in Section III-C. Finally, the SDR approach is adopted to analyze the upper-bound performance, and is presented in Section III-D.
\subsection{Problem Formulation}
As demonstrated in \cite{kalantari2016directional}, the study of the DM technique can be based on strict phase and relaxed phase constraints. For the strict phase-based waveform design, the received signal $y_k$ should have exactly the same phase as the induced symbol of the $k$-th CU (i.e., $s_k$), which constrains the degrees of freedom (DoFs) in designing the waveform ${\mathbf{x}}$. Hence, inspired by the concept of CI \cite{masouros2015exploiting,masouros2009dynamic}, the optimization problem is proposed to locate the received symbol for each CU within a constructive region rather than restrict the symbol in the proximity of the constellation point, namely the relaxed phase based design.
\\\indent The CI technique has been widely investigated in the recent work. To avoid deviating our focus, we will omit the derivation of the CI constraints, and refer the reader to \cite{masouros2015exploiting} for more details. Since CI-based waveform design aims to transform the undesirable MUI into useful power by pushing the received signal further away from the $M$-PSK decision boundaries, all interference contributes to the useful received power \cite{xu2020rethinking}. Herewith, the SNR of the $k$-th user is expressed as
\begin{equation}\label{eq9}
    {\text{SN}}{{\text{R}}_k} = \frac{{{{\left| {{\mathbf{h}}_k^H{\mathbf{x}}} \right|}^2}}}{{\sigma _{{C_k}}^2}}.
\end{equation}
\\\indent With the knowledge of the channel information, all CUs' data, as well as the location of target and clutter resources is readily available at the transmitter, we formulate the following optimization problem aiming at maximizing the SINR of the target return
\begin{equation}\label{eq10}
\begin{gathered}
  \mathop {\max }\limits_{{\mathbf{w}},{\mathbf{x}}} \;\;\;\;{\text{  SIN}}{{\text{R}}_{rad}} \hfill \\
  \text{s.t.}\;\;\;\;\;{\text{    }}{\left\| {\mathbf{x}} \right\|^2} \leq {P_0} \hfill \\
  \;\;\;\;\;\;\;\; {\text{         }}\left| {\arg \left( {{\mathbf{h}}_k^H{\mathbf{x}}} \right) - \arg \left( {{s_k}} \right)} \right| \leq \xi  , \forall k, \hfill \\
  \;\;\;\;\;\;\;\;\;\;  {\text{SN}}{{\text{R}}_k} \geq {\Gamma _k},  \forall k, \hfill \\
\end{gathered}
\end{equation}
where $P_0$ denotes the transmit power budget, $\Gamma_k$ is the given SNR threshold, and $\xi$ is the phase threshold where the noise-less received symbols are supposed to lie.
\\\indent As illustrated in Fig. 2, by taking one of the QPSK constellation points as an example, the constructive region is given as the green area. In Fig. 2(a), ${{\bar y}_k}$ denotes the noise-excluding signal and the SNR related scalar $\gamma_k$ is the threshold distance to the decision region of the received symbol at the $k$-th CU. Then, in order to express the constructive region geometrically, we rotate the noise-free received signal ${{\bar y}_k}$ and project it onto real and imaginary axes, which is illustrated in Fig. 2(b). By noting $\left| {{s_k}} \right| = 1$, the rotated signal $\tilde y_k$ can be given in the form of
\begin{equation}\label{eq11}
\begin{aligned}
    {{\tilde y}_k} = \left( {{y_k} - {n_k}} \right)\frac{{s_k^*}}{{\left| {{s_k}} \right|}}
    &= {\mathbf{h}}_k^H{\mathbf{x}}s_k^*\\ & = {\mathbf{\tilde h}}_k^H{\mathbf{x}},
\end{aligned}
\end{equation}
where ${{{\mathbf{\tilde h}}}_k} = {{\mathbf{h}}_k}s_k^*$. Let us represent $\operatorname{Re} \left( {{{\tilde y}_k}} \right) = \operatorname{Re} \left( {{\mathbf{\tilde h}}_k^H{\mathbf{x}}} \right)$ and $\operatorname{Im} \left( {{{\tilde y}_k}} \right) = \operatorname{Im} \left( {{\mathbf{\tilde h}}_k^H{\mathbf{x}}} \right)$. Then, the ${\text{SINR}_{rad}}$ maximization problem (\ref{eq9}) can be recast as \cite{masouros2015exploiting}
\begin{subequations}\label{eq12}
\begin{align}
  &\mathop {\max }\limits_{{\mathbf{w}},{\mathbf{x}}} {\text{  SIN}}{{\text{R}}_{rad}} \hfill \\
  &\text{s.t.}\;\;\;\;\; {\left\| {\mathbf{x}} \right\|^2} \leq {P_0} \hfill \\
  &\;\;\;\;\;\;\;\;\; \left| {\operatorname{Im} \left( {{\mathbf{\tilde h}}_k^H{\mathbf{x}}} \right)} \right| \leq \left( {\operatorname{Re} \left( {{\mathbf{\tilde h}}_k^H{\mathbf{x}}} \right) - \sqrt {\sigma _{{C_k}}^2{\Gamma _k}} } \right)\tan \phi, \forall k
\end{align}
\end{subequations}
where $\phi {\text{ = }} \pm {\pi  \mathord{\left/{\vphantom {\pi  M}} \right. \kern-\nulldelimiterspace} M}$.
\subsection{Solve (12) by SQ Approach}
It is noted that problem (12) is still non-convex since the clutter is signal-dependent, where the quadratic form of optimizing variable $\mathbf{x}$ is included in both numerator and denominator. To address this issue, in this section we develop an SQ approach to extract a suboptimal solution. Firstly, note that problem (12) can be viewed as the classical minimum variance distortionless response (MVDR) beamforming problem with respect to $\mathbf{w}$, which can be expressed as a function of $\mathbf{x}$ as
\begin{equation}\label{eq13}
    {\mathbf{w}} = \frac{{{{\left[ {{\mathbf{\Sigma }}\left( {\mathbf{x}} \right) + {\mathbf{I}}} \right]}^{ - 1}}{\mathbf{U}}\left( {{\theta _0}} \right){\mathbf{x}}}}{{{{\mathbf{x}}^H}{{\mathbf{U}}^H}\left( {{\theta _0}} \right){{\left[ {{\mathbf{\Sigma }}\left( {\mathbf{x}} \right) + {\mathbf{I}}} \right]}^{ - 1}}{\mathbf{U}}\left( {{\theta _0}} \right){\mathbf{x}}}}.
\end{equation}
By substituting (13) into (4), the optimization problem (12) can be rewritten as \cite{7450660,cui2013mimo}
\begin{equation}\label{eq14}
\begin{aligned}
    &\mathop {\max }\limits_{{\mathbf{x}}} {\text{  }}{{\mathbf{x}}^H}{\mathbf{\Phi }}\left( {\mathbf{x}} \right){\mathbf{x}} \hfill \\
    &\text{s.t.}\;12\left( b \right)\;{\text{and}}\;12\left( c \right),
\end{aligned}
\end{equation}
where ${\mathbf{\Phi }}\left( {\mathbf{x}} \right) = {\mathbf{U}}^H{\left( {{\theta _0}} \right)}{\left[ {{\mathbf{\Sigma }}\left( {\mathbf{x}} \right) + {\mathbf{I}}} \right]^{ - 1}}{\mathbf{U}}\left( {{\theta _0}} \right)$ is a positive-semidefinite SINR matrix. To solve problem (14), we adopt the sequential optimization algorithm (SOA) presented in \cite{cui2013mimo}. To be specific, let us firstly ignore the dependence of ${\mathbf{\Phi }}\left( {\mathbf{x}} \right)$ on $\mathbf{x}$, i.e., fixing the signal-dependent matrix ${\mathbf{\Phi }}\left( {\mathbf{x}} \right)={\mathbf{\Phi }}$ for a given ${\mathbf{x}}$. To start with, we initialize ${\mathbf{\Phi }}={\mathbf{\Phi }}_0$, where ${\mathbf{\Phi }}_0$ is a constant positive-semidefinite matrix. By using SOA, the waveform ${\mathbf{x}}$ is optimized iteratively with the updated ${\mathbf{\Phi}}$ till convergence. By doing so, in each SOA iteration we solve the following problem
\begin{equation}\label{eq15}
\begin{aligned}
    &\mathop {\max }\limits_{{\mathbf{x}}} {\text{  }}{{\mathbf{x}}^H}{\mathbf{\Phi }}{\mathbf{x}}\hfill \\
    &\text{s.t.}\;12\left( b \right)\;{\text{and}}\;12\left( c \right).
\end{aligned}
\end{equation}
Note that problem (15) is easily converted to a convex Quadratically Constrained
Quadratic Program (QCQP) problem by recasting the signal-independent matrix ${\mathbf{\Phi }}$ to be negative-semidefinite as follows \cite{7450660}
\begin{equation}\label{eq16}
\begin{aligned}
    &\mathop {\max }\limits_{\mathbf{x}}{\text{  }}{{\mathbf{x}}^H}{\mathbf{Qx}} \hfill \\
    &\text{s.t.}\;12\left( b \right)\;{\text{and}}\;12\left( c \right),
\end{aligned}
\end{equation}
where ${\mathbf{Q}} = \left( {{\mathbf{\Phi }} - \lambda {\mathbf{I}}} \right)$,  $\lambda  \ge {\lambda _{\max }}\left( {\mathbf{\Phi }} \right)$, where ${\lambda _{\max }}\left( {\mathbf{\Phi }} \right)$ is the largest eigenvalue of $\mathbf{\Phi}$. It is straightforward to see that $\mathbf{Q}$ is negative-semidefinite, thus the objective function is concave, and then it can be tackled efficiently by CVX toolbox \cite{grant2014cvx}. Here, we denote $\mathbf{w^*}$ and $\mathbf{x^*}$ as the optimal receive beamformer and waveform, respectively. Furthermore, as the expression given in (13), the receive beamforming vector $\mathbf{w^*}$ can be updated by the optimal waveform $\mathbf{x^*}$. Therefore, the suboptimal solutions are obtained until convergence by updating $\mathbf{x}$ and $\mathbf{w}$ iteratively. The generated solution will serve as a baseline in Section VI named as SQ. For clarity, we summarize the SQ approach in Algorithm 1.
\\\indent In SQ approach, we note that the reformulation of the objective function in (16) actually relaxes the one given in (15). To be specific, we have ${\mathbf{x}^{H}}{\mathbf{Qx}} = {\mathbf{x}^H}\left({{\mathbf{\Phi }} - \lambda {\mathbf{I}}} \right){\mathbf{x}} = {\mathbf{x}^{H}}{\mathbf{\Phi x}}-\lambda {\mathbf{x}^H}{\mathbf{x}}$, while the power constraint (12b) indicates that ${\mathbf{x}^H}{\mathbf{x}}$ in the second term is not constant. In the following subsection, we adopt FP algorithm to solve problem (15), which aims to tackle the problem without a relaxation in the objective function.
\renewcommand{\algorithmicrequire}{\textbf{Input:}}
\renewcommand{\algorithmicensure}{\textbf{Output:}}
\begin{algorithm}
\caption{SQ Algorithm for solving problem (12)}
\label{alg:1}
\begin{algorithmic}
    \REQUIRE ${P_0},{\mathbf{h}}_k,{{\sigma _{{C_k}}^2}},{\sigma _R^2},\theta_i,\theta_0,\alpha_0, b_i, \Gamma_k, \forall k,\forall i, \varepsilon > 0$, and the maximum iteration number $m_{max}$
    \ENSURE ${\mathbf{x}}$
    \STATE 1. Reformulate problem (12) by (16).
    \STATE 2. Initialize the positive-semidefinite matrix ${\mathbf{\Phi}}^0$, $m = 1$.
    \WHILE {$m \le {m_{max}}$ and $\left| \text{SINR}_{rad}^m - \text{SINR}_{rad}^{m-1} \right| \ge \varepsilon$ }
    \STATE 3. Calculate $\mathbf{Q}^{m-1}$, solve problem (16) to obtain the optimal waveform ${{\mathbf{x}}^m}$.
    \STATE 4. Update $\mathbf{\Phi}^m$ by ${{\mathbf{x}}^m}$.
    \STATE 5. Transform $\mathbf{\Phi}^m$ into the negative-semidefinite matrix $\mathbf{Q}^m$.
    \STATE 6. $ m = m + 1$.
    \ENDWHILE
\end{algorithmic}
\end{algorithm}
\subsection{Solve (12) by FP Approach}
The original radar SINR maximization problem can also be written as
\begin{equation}\label{eq17}
\begin{gathered}
  \mathop {\max }\limits_{\mathbf{x}} {\text{  }}\frac{{\mu {{\left| {{{\mathbf{w}}^H}{\mathbf{U}}\left( {{\theta _0}} \right){\mathbf{x}}} \right|}^2}}}{{{{\mathbf{w}}^H}\left( {{\mathbf{\Sigma }}\left( {\mathbf{x}} \right) + {{\mathbf{I}}_{{N_R}}}} \right){\mathbf{w}}}} \hfill \\
  \text{s.t.}\;12\left( b \right)\;{\text{and}}\;12\left( c \right). \hfill \\
\end{gathered}
\end{equation}
We note that the non-convexity lies only in the objective function in the problem above, and one can stay in the convex feasible region by exploiting various linear iteration schemes. Thus, it can be solved by converting the objective function into its linear approximation form. Following the  \textit{Dinkelbach's transform} of FP problem presented in \cite{shen2018fractional}, we firstly reformulate the objective function as
\begin{equation}\label{eq18}
\begin{gathered}
  \mathop {\max }\limits_{\mathbf{x}} {\text{   }}\mu {\left| {{{\mathbf{w}}^H}{\mathbf{U}}\left( {{\theta _0}} \right){\mathbf{x}}} \right|^2} - u{{\mathbf{w}}^H}\left( {{\mathbf{\Sigma }}\left( {\mathbf{x}} \right) + {{\mathbf{I}}_{{N_R}}}} \right){\mathbf{w}} \hfill \\
  \text{s.t.}\;12\left( b \right)\;{\text{and}}\;12\left( c \right). \hfill \\
\end{gathered}
\end{equation}
Here, the objective function is still non-concave because of the first term. To proceed with optimization problem (18), let us firstly denote $f\left( {\mathbf{x}} \right) = {\left| {{{\mathbf{w}}^H}{\mathbf{U}}\left( {{\theta _0}} \right){\mathbf{x}}} \right|^2}$. Then, we approximate the objective function $f\left( {\mathbf{x}} \right)$ by its first-order Taylor expansion with respective to $\mathbf{x}$ at ${\mathbf{x}}' \in \mathcal{D}$, where $\mathcal{D}$ denotes the feasible region of (17).
\begin{equation}\label{eq19}
\begin{aligned}
  f\left( {\mathbf{x}} \right) &\approx f\left( {{\mathbf{x'}}} \right){ + }\nabla {f^H}\left( {{\mathbf{x'}}} \right)\left( {{\mathbf{x}} - {\mathbf{x'}}} \right) \hfill \\
   & = f\left( {{\mathbf{x'}}} \right) + \hfill\\ &\operatorname{Re} \left( {{{\left( {2\left( {{{{\mathbf{x'}}}^H}{{\mathbf{U}}^H}\left( {{\theta _0}} \right){\mathbf{w}}} \right){{\mathbf{U}}^H}\left( {{\theta _0}} \right){\mathbf{w}}} \right)}^H}\left( {{\mathbf{x}} - {\mathbf{x'}}} \right)} \right), \hfill \\
\end{aligned}
\end{equation}
where $\nabla f\left(  \cdot  \right)$ denotes the gradient of $f\left(  \cdot  \right)$. For simplicity, we omit the constant term $f\left( {{\mathbf{x'}}} \right)$ and denote
\begin{equation}\label{eq20}
\begin{aligned}
    &g\left( {\mathbf{x}} \right) = \hfill \\
    &\operatorname{Re} \left( {{{\left( {2\left( {{{\mathbf{x}}^{ {m - 1} }}^{^H}{{\mathbf{U}}^H}\left( {{\theta _0}} \right){\mathbf{w}}} \right){{\mathbf{U}}^H}\left( {{\theta _0}} \right){\mathbf{w}}} \right)}^H}\left( {{\mathbf{x}} - {{\mathbf{x}}^{ {m - 1} }}} \right)} \right).
\end{aligned}
\end{equation}
Herewith, the $m$-th iteration of the FP algorithm can be obtained by solving the following convex optimization problem
\begin{equation}\label{eq21}
\begin{gathered}
  \mathop {\max }\limits_{\mathbf{x}} {\text{   }}\mu g\left( {\mathbf{x}} \right) - u{{\mathbf{w}}^H}\left( {{\mathbf{\Sigma }}\left( {\mathbf{x}} \right) + {{\mathbf{I}}_{{N_R}}}} \right){\mathbf{w}} \hfill \\
  \text{s.t.}\;12\left( b \right)\;{\text{and}}\;12\left( c \right), \hfill \\
\end{gathered}
\end{equation}
where ${{\mathbf{x}}^{m-1}} \in \mathcal{D}$ is the point obtained at the $\left(m-1\right)$-th iteration. The optimal solution ${{\mathbf{x}}^ m } \in \mathcal{D}$ can be obtained by solving problem (21), and then the receive beamformer ${\mathbf{w}^m}$ can be obtained by substituting ${{\mathbf{x}}^ m }$ in (13). Furthermore, $u$ is an auxiliary variable, which is updated iteratively by
\begin{equation}\label{eq22}
    {u^{m + 1}} = \frac{{\mu {{\left| {{{\mathbf{w}}^H}{\mathbf{U}}\left( {{\theta _0}} \right){{\mathbf{x}}^m}} \right|}^2}}}{{{{\mathbf{w}}^H}\left( {{\mathbf{\Sigma }}\left( {{{\mathbf{x}}^m}} \right) + {{\mathbf{I}}_{{N_R}}}} \right){\mathbf{w}}}}.
\end{equation}
It is easy to prove the convergence of the algorithm given the non-increasing property of $y$ during each iteration \cite{shen2018fractional}. For clarity, we summarize the above in Algorithm 2. We note that the computational complexity of solving problem (21) at each iteration is given by $\mathcal{O} \left( N_T^3\sqrt {K + 1}  \right)$ \cite{nesterov1994interior}.
\subsection{Upper Bound Performance}
In this subsection, we derive a new optimization problem to analyze the upper bound performance of problem (12). According to the reformulation given in problem (14), the objective function is equivalent to
\begin{equation}\label{eq23}
    y\left( \mathbf{x}\right) = {{\mathbf{x}}^H}{{\mathbf{U}}^H}\left( {{\theta _0}} \right){\left[ {{\mathbf{\Sigma }}\left( {\mathbf{x}} \right) + {\mathbf{I}}} \right]^{ - 1}}{\mathbf{U}}\left( {{\theta _0}} \right){\mathbf{x}}.
\end{equation}
It is obvious that ${{{\mathbf{\Sigma }}\left( {\mathbf{x}} \right) + {\mathbf{I}}} } \succeq {\mathbf{I}}$, and thereby, $\left[ {{\mathbf{\Sigma }}\left( {\mathbf{x}} \right) + {\mathbf{I}}} \right]^{ - 1} \preceq  {\mathbf{I}}$, which indicates that  $y\left( \mathbf{x}\right) \le {{\mathbf{x}}^H}{{\mathbf{U}}^H}\left( {{\theta _0}} \right){\mathbf{U}}\left( {{\theta _0}} \right){\mathbf{x}}$. So we firstly relax the objective function as
\begin{equation}\label{eq24}
\begin{aligned}
  &\mathop {\max }\limits_{\mathbf{x}} {\text{ }}{{\mathbf{x}}^H}{{\mathbf{U}}^H}\left( {{\theta _0}} \right){\mathbf{U}}\left( {{\theta _0}} \right){\mathbf{x}} \hfill \\
  &\text{s.t.}\;12\left( b \right)\;{\text{and}}\;12\left( c \right).
\end{aligned}
\end{equation}
It is noted that problem (24) is an inhomogeneous QCQP \cite{vandenberghe1996semidefinite} problem. We firstly define ${\mathbf{X}} = {\mathbf{x}}{{\mathbf{x}}^H}$ and let
\begin{equation}\label{eq25}
    {\mathbf{\tilde X}} = \left[ {\begin{array}{*{20}{c}}
    {\mathbf{X}}&{\mathbf{x}} \\
    {{{\mathbf{x}}^H}}&1
\end{array}} \right].
\end{equation}
Afterwards, problem (24) can be recast as
\begin{equation}\label{eq26}
\begin{gathered}
  \mathop {\max }\limits_{\mathbf{x},\mathbf{X}} {\text{ tr}}\left( {{\mathbf{X}}{{{\mathbf{\hat U}}}_0}} \right) \hfill \\
  \text{s.t.} \;\; {\mathbf{\tilde X}} \succeq 0, {\text{rank}}\left( {{\mathbf{\tilde X}}} \right) = 1 \hfill \\
   \;12(b)\;\;\text{and}\;\;12(c),
\end{gathered}
\end{equation}
where ${{{\mathbf{\hat U}}}_0}{\text{ = }}{{\mathbf{U}}^H}\left( {{\theta _0}} \right){\mathbf{U}}\left( {{\theta _0}} \right)$. Note that problem (26) is readily to be solved by the SDR technique \cite{park2017general}. To start with, we relax the above optimization problem by dropping the rank-1 constraint, yielding
\begin{equation}\label{eq27}
\begin{gathered}
  \mathop {\max }\limits_{\mathbf{x},\mathbf{X}} {\text{ tr}}\left( {{\mathbf{X}}{{{\mathbf{\hat U}}}_0}} \right) \hfill \\
  \text{s.t.} \;\;\;  {\mathbf{\tilde X}} \succeq 0 \hfill \\
  \;\;\;\;\;\;12(b)\;\;\text{and}\;\;12(c).
\end{gathered}
\end{equation}
Problem (27) is convex and can be optimally solved. Here, we define ${{\mathbf{X}}^*}$ and ${{\mathbf{x}}^*}$ as the approximate solution to the problem above. By substituting the ${{\mathbf{X}}^*}$ in the objective function in (25), the optimal objective value is an upper bound of the optimal value in problem (12).
\begin{remark}\label{rmk:1}
    In problem (27), the constraint ${\mathbf{\tilde X}} \succeq 0$ implies ${\mathbf{X}} \succeq {\mathbf{x}}{{\mathbf{x}}^H}$. Based on the relaxations above, we have the following inequalities
    \begin{equation}
        {\mathrm{tr}}\left( {{{\mathbf{X}}^*}{{{\mathbf{\hat U}}}_0}} \right) \geq {\mathrm{tr}}\left( {{{\mathbf{x}}^*}{{\mathbf{x}}^*}^H{{{\mathbf{\hat U}}}_0}} \right) \geq {{\mathbf{x}}^*}^H{\mathbf{\Phi }}\left( {{{\mathbf{x}}^*}} \right){{\mathbf{x}}^*} \nonumber
    \end{equation}
    Therefore, the objective value in (27) is larger than the achievable $\text{SINR}_{rad}$, of which performance is presented as the upper bound of radar receive SINR in our simulation results.
\end{remark}
\renewcommand{\algorithmicrequire}{\textbf{Input:}}
\renewcommand{\algorithmicensure}{\textbf{Output:}}
\begin{algorithm}
\caption{The Proposed FP Algorithm for solving problem (12)}
\label{alg:2}
\begin{algorithmic}
    \REQUIRE ${P_0},{\mathbf{h}}_k,{{\sigma _{{C_k}}^2}},{\sigma _R^2},\theta_i,\theta_0,\alpha_0, b_i, \Gamma_k, \forall k,\forall i, \varepsilon > 0$, and the maximum iteration number $m_{max}$
    \ENSURE ${\mathbf{x}}$
    \STATE 1. Reformulate the objective function as given in (21).
    \STATE 2. Initialize ${\mathbf{x}}^0 \in\ \mathcal{D}$ randomly, $m = 1$.
    \WHILE {$m \le {m_{max}}$ and $\left| \text{SINR}_{rad}^m - \text{SINR}_{rad}^{m-1} \right| \ge \varepsilon$ }
    \STATE 3. Solve problem (21) to obtain the optimal waveform ${{\mathbf{x}}^m}$.
    \STATE 4. Obtain the receive beamformer ${{\mathbf{w}}^m}$ by substituting ${{\mathbf{x}}^m}$ in (13).
    \STATE 5. Update $u$ by (22).
    \STATE 6. $ m = m + 1$.
    \ENDWHILE
\end{algorithmic}
\end{algorithm}
\section{${\text{SIN}}{{\text{R}}_{rad}}$ Maximization With Target Location Uncertainty}
In a practical target tracking scenario, the target location is not perfectly known to the BS due to its movement and random fluctuation, and we therefore consider the scenario where a rough estimation of the target's angle is available at the BS. That is, the target is assumed to locate in an uncertain angular interval. In the following waveform design, we aim to maximize the minimum ${\text{SINR}_{rad}}$ with regard to all possible locations within the interval, while taking CI technique and power budget into account. Finally, an efficient solver is proposed to tackle the worst-case optimization problem.
\subsection{Problem Formulation}
Let us denote the uncertain interval as $\Psi = \left[\theta_0-\Delta\theta, \theta_0+\Delta\theta \right]$. It is noteworthy that the target from every possible direction should be taken into account when formulating the optimization problem. To this end, we therefore consider the following worst-case problem, which is to maximize the minimum ${\text{SINR}_{rad}}$ with respect to all the possible target locations within $\Psi$. For the sake of simplicity, let $\theta_p \in {\text{card}} \left(\Psi\right)$ denote the $p$-th possible location in the given region, where $ {\text{card}} \left(\cdot \right)$ represents the cardinality of $\left(  \cdot  \right)$.
\begin{equation}\label{eq28}
\begin{aligned}
  &\mathop {\max }\limits_{\mathbf{x}} {\text{  }}\mathop {{\text{min}}}\limits_{{\theta _p} \in {\text{card}}\left( \Psi  \right)}\frac{{\mu {{\left| {{{\mathbf{w}}^H}{\mathbf{U}}\left( {{\theta _p}} \right){\mathbf{x}}} \right|}^2}}}{{{{\mathbf{w}}^H}\left( {{\mathbf{\Sigma }}\left( {\mathbf{x}} \right) + {{\mathbf{I}}_{{N_R}}}} \right){\mathbf{w}}}} \hfill \\
   & \text{s.t.}\;12\left( b \right)\;{\text{and}}\;12\left( c \right). \hfill \\
\end{aligned}
\end{equation}
Note that the problem above is non-convex since the point-wise maximum of concave functions is not convex. In the following subsection, we will work on solving the problem (28).
\subsection{Efficient Solver}
As is detailed in \cite{shen2018fractional}, the straightforward extension of Dinkelbach's transform which is deployed in Section III does not guarantee the equivalence to problem (28). Thus, we give the equivalent quadratic transformation of the \textit{the max-min-ratio} problem (28), which is rewritten as
\begin{equation}\label{eq29}
\begin{gathered}
  \mathop {\max }\limits_{{\mathbf{x}},{\mathbf{u}}} {\text{  }}\mathop {{\text{min}}}\limits_{{\beta _p} \in {\text{card}}\left( \Psi  \right)} {\text{ }}2{u_p}\sqrt {\mu {{\left| {{{\mathbf{w}}^H}{\mathbf{U}}\left( {{\theta _p}} \right){\mathbf{x}}} \right|}^2}}  \hfill\\
  \;\;\;\;\;\;\;\;\;\;\;\;\;\;\;\;\;\;\;\;\;\;\;\;\; - u_p^2{{\mathbf{w}}^H}\left( {{\mathbf{\Sigma }}\left( {\mathbf{x}} \right) + {{\mathbf{I}}_{{N_R}}}} \right){\mathbf{w}} \hfill \\
  \text{s.t.}\;12\left( b \right)\;{\text{and}}\;12\left( c \right). \hfill \\
\end{gathered}
\end{equation}
Here, we denote $\mathbf{u}$ as a collection of variables $\left\{ {{u_1}, \cdots ,{u_P}} \right\}, {u_p} \in \mathbb{R}$. The objective above is a sequence of ratios for $\theta_p \in {\text{card}} \left(\Psi\right)$. To proceed\footnote{As given in the expression (4), it can be found that the objective function is independent with the amplitude coefficient $\alpha_0$, therefore, when the target location is imperfectly known, the uncertainty of amplitude can be neglected in the problem formulation.},  we rewrite problem (29) in an epigraph form by introducing the variable $a, {a} \in \mathbb{R}$, which yields the following formulation
\begin{subequations}\label{eq30}
\begin{align}
  &\mathop {\max }\limits_{{\mathbf{x}},{\mathbf{u}},a} {\text{  }}\;\;\;a \hfill \\
  &\text{s.t.}{\text{  }}2{u_p}\sqrt {\mu {{\left| {{{\mathbf{w}}^H}{\mathbf{U}}\left( {{\theta _p}} \right){\mathbf{x}}} \right|}^2}}  - u_p^2{{\mathbf{w}}^H}\left( {{\mathbf{\Sigma }}\left( {\mathbf{x}} \right) + {{\mathbf{I}}_{{N_R}}}} \right){\mathbf{w}} \ge a, \nonumber\\
  &\;\;\;\;\;\;\;\;\;\;\;\;\;\;\;\;\;\;\;\;\;\;\;\;\;\;\;\;\;\;\;\;\;\;\;\;\;\;\;\;\;\;\;\;\;\;\;\;\;\;\;\;\;\;\;\;\;\forall {{\theta _p} \in {\text{card}}\left( \Psi  \right)} \hfill \\
  & \;\;\;\;\;12\left( b \right){\text{  and  }}12\left( c \right).
\end{align}
\end{subequations}
By observing problem (30), it is noted that the constraint (30b) is non-convex. To tackle the problem, likewise, we substitute ${\mu {{\left| {{{\mathbf{w}}^H}{\mathbf{U}}\left( {{\theta _p}} \right){\mathbf{x}}} \right|}^2}}$ in the first term of (30b) with its first-order Taylor expansion approximation with respective to $\mathbf{x}$ at ${\mathbf{x}}' \in \mathcal{D}$ as is given in (19), which is expressed as
\begin{equation}\label{eq31}
\begin{aligned}
  &\mathop {\max }\limits_{{\mathbf{x}},{\mathbf{u}},a} {\text{  }}\;\;\;a \hfill \\
  &\text{s.t.}{\text{  }}2{u_p}\sqrt {\mu\operatorname{Re} \left( {{{\left( {2\left( {{{{\mathbf{x'}}}^H}{{\mathbf{U}}^H}\left( {{\theta _p}} \right){\mathbf{w}}} \right){{\mathbf{U}}^H}\left( {{\theta _0}} \right){\mathbf{w}}} \right)}^H}\left( {{\mathbf{x}} - {\mathbf{x'}}} \right)} \right)}  -  \hfill\\ &\;\;\;\;\;\;\;\;\;\;\;\;\;\;\;u_p^2{{\mathbf{w}}^H}\left( {{\mathbf{\Sigma }}\left( {\mathbf{x}} \right) + {{\mathbf{I}}_{{N_R}}}} \right){\mathbf{w}} \ge a, \;\;\forall {{\theta _p} \in {\text{card}}\left( \Psi  \right)} \hfill \\
  & \;\;\;\;\;12\left( b \right){\text{  and  }}12\left( c \right).
\end{aligned}
\end{equation}
It is noted that at the $m$-th iteration, ${\mathbf{x}}'$ in problem (31) denotes ${{\mathbf{x}}^{m-1}} \in \mathcal{D}$, which is the point obtained at the $\left(m-1\right)$-th iteration. When the optimal waveform $\mathbf{x}$ is obtained, the variable $u_p$ can be updated by the following closed form as
\begin{equation}\label{eq32}
    {u_p^{m+1}} = \frac{{\sqrt {\mu {{\left| {{{\mathbf{w}}^H}{\mathbf{U}}\left( {{\theta _p}} \right){\mathbf{x}^m}} \right|}^2}} }}{{{{\mathbf{w}}^H}\left( {{\mathbf{\Sigma }}\left( {\mathbf{x}^m} \right) + {{\mathbf{I}}_{{N_R}}}} \right){\mathbf{w}}}}.
\end{equation}
Now, problem (31) can be solved by interior point methods at a worst-case computational complexity of $\mathcal{O} \left( N_T^3\sqrt {{\Psi _0} + K + 1} \right)$ at each iteration\cite{nesterov1994interior}, where we denote $ \Psi_0 $ as the number of elements in $ {\text{card}} \left(\Psi\right) $. For clarity, the proposed method of solving (28) is summarized in Algorithm 3.
\renewcommand{\algorithmicrequire}{\textbf{Input:}}
\renewcommand{\algorithmicensure}{\textbf{Output:}}
\begin{algorithm}
\caption{The Proposed Algorithm for solving multiple-ratio FP problem (28)}
\label{alg:3}
\begin{algorithmic}
    \REQUIRE ${P_0},{\mathbf{h}}_k,{{\sigma _{{C_k}}^2}},{\sigma _R^2},\theta_i,\theta_0,\alpha_0, b_i, \Gamma_k, \Delta\theta, \forall k,\forall i, \varepsilon > 0$, and the maximum iteration number $m_{max}$
    \ENSURE ${\mathbf{x}}$
    \STATE 1. Reformulate the problem by (29).
    \STATE 2. Transform the problem to epigraph form following (30).
    \STATE 3. Reformulate the non-convex constraint by (31).
    \STATE 4. Initialize ${\mathbf{x}}^0 \in\ \mathcal{D}$ randomly, $m = 1$.
    \WHILE {$m \le {m_{max}}$ and $|| \mathbf{u}^m - \mathbf{u}^{m-1} || \ge \varepsilon $ }
    \STATE 5. Solve problem (31) to obtain the optimal waveform ${{\mathbf{x}}^m}$.
    \STATE 6. Obtain the receive beamformer ${{\mathbf{w}}^m}$ by substituting ${{\mathbf{x}}^m}$ in (13).
    \STATE 7. Update $\mathbf{u}$ by (32).
    \STATE 8. $ m = m + 1$.
    \ENDWHILE
\end{algorithmic}
\end{algorithm}
\section{CI Precoding with Destructive Interference to the Radar Receiver}
In this section, we consider the information transmission security of the DFRC system. We assume that the communication users are legitimate, and treat the point-like target as a potential eavesdropper which might surveille the information from BS to CUs. Accordingly, in the following design, we aim to maximize the SINR at radar receiver like the proposed formulation in Section III and Section IV, while confining the received signal at the target into the destructive region of the constellation, in order to ensure the PHY security for DFRC transmission. This problem will be studied under the circumstances that target location is known to the BS perfectly and imperfectly, respectively.
\subsection{With Knowledge of Precise Target Location}
In prior work with respect to DM technique, such as algorithms proposed in \cite{kalantari2016directional}, the problems are designed based on the CSI of legitimate users, where the symbols received by potential eavesdroppers are scrambled due to the channel disparity. However, PHY security cannot be explicitly guaranteed in this way. To be specific, Taking QPSK modulation as an example, the intended symbol can be intercepted with a $\frac{1}{4}$ probability at the target when the target's channel is independent with the CUs' channels, while more importantly, the probability of the target intercepting increases when the target and CUs' channels are correlated. The simulation result will be shown in Section VI.
\\\indent While the CI-based precoding guarantees low SER at CUs, we still need to focus on the detection performance at the target in order to prevent the transmit information from being decoded. Thus, the following problem is designed to improve the SER at the target. In detail, we define the region out of the constructive region as destructive region and aims at restricting the received signal of the potential eavesdropper in the destructive area.
\\\indent We firstly take $s_1$ as a reference. Likewise, the received noise-excluding signal at the target can be expressed as
\begin{equation}\label{eq33}
\begin{aligned}
  {{\tilde y}_R} = \left( {{y_R} - e} \right)\frac{{s_1^ * }}{{\left| {{s_1}} \right|}}
  &= {\alpha _0}{\mathbf{a}}_t^H\left( \theta_0  \right){\mathbf{x}}s_1^ *  \hfill\\
   &= {\alpha _0}{\mathbf{\tilde a}}_t^H\left( \theta_0  \right){\mathbf{x}},
\end{aligned}
\end{equation}
where ${\mathbf{\tilde a}}_t^H\left( \theta_0  \right) = {\mathbf{a}}_t^H\left( \theta_0  \right)s_1^ * $. Accordingly, the destructive region can be described by
\begin{equation}\label{eq34}
    \left| {\operatorname{Im} \left( {{{\tilde y}_R}} \right)} \right| \geq \left( {\operatorname{Re} \left( {{{\tilde y}_R}} \right) - \sqrt {\sigma _T^2{\Gamma _T}} } \right)\tan \phi.
\end{equation}
where the scalar $\Gamma_T$ denotes the desired maximum SNR for the potential eavesdropper and $\sqrt {\sigma _T^2{\Gamma _T}}$ corresponds to $\gamma_e$ in Fig. 3. As illustrated in Fig. 3, the destructive region can be divided to three zones and the inequality (34) holds when any one of the following constraints is fulfilled.
\begin{subequations}\label{eq35}
\begin{align}
    &zone\;1: \operatorname{Re} \left( {{\alpha _0}{\mathbf{\tilde a}}_t^H\left( \theta_0  \right){\mathbf{x}}} \right) - \sqrt {\sigma _T^2{\Gamma _T}}  \leq 0 \hfill \\
    &zone\;2: \nonumber \\
    &\operatorname{Im} \left( {{\alpha _0}{\mathbf{\tilde a}}_t^H\left( \theta_0  \right){\mathbf{x}}} \right) \geq \left( {\operatorname{Re} \left( {{\alpha _0}{\mathbf{\tilde a}}_t^H\left( \theta_0  \right){\mathbf{x}}} \right) - \sqrt {\sigma _T^2{\Gamma _T}} } \right)\tan \phi \nonumber \\
    &{\text{and}}\; \operatorname{Re} \left( {{\alpha _0}{\mathbf{\tilde a}}_t^H\left( \theta_0  \right){\mathbf{x}}} \right) > \sqrt {\sigma _T^2{\Gamma _T}} \\
    &zone\;3: \nonumber \\
    &- \operatorname{Im} \left( {{\alpha _0}{\mathbf{\tilde a}}_t^H\left( \theta_0  \right){\mathbf{x}}} \right) \geq \left( {\operatorname{Re} \left( {{\alpha _0}{\mathbf{\tilde a}}_t^H\left( \theta_0  \right){\mathbf{x}}} \right) - \sqrt {\sigma _T^2{\Gamma _T}} } \right)\tan \phi \nonumber \\
    &{\text{and}}\; \operatorname{Re} \left( {{\alpha _0}{\mathbf{\tilde a}}_t^H\left( \theta_0  \right){\mathbf{x}}} \right) > \sqrt {\sigma _T^2{\Gamma _T}}.
\end{align}
\end{subequations}
For simplicity, we denote (35) as destructive interference (DI) constraints. By taking the full region of destructive interference into consideration, the optimization problem can be formulated as
\begin{equation}\label{eq36}
\begin{aligned}
    & \mathop {\max }\limits_{\mathbf{x}} {\text{  }}\frac{{\mu {{\left| {{{\mathbf{w}}^H}{\mathbf{U}}\left( {{\theta _0}} \right){\mathbf{x}}} \right|}^2}}}{{{{\mathbf{w}}^H}\left( {{\mathbf{\Sigma }}\left( {\mathbf{x}} \right) + {{\mathbf{I}}_{{N_R}}}} \right){\mathbf{w}}}} \hfill \\
    &\text{s.t.}\;12\left( b \right)\;{\text{and}}\;12\left( c \right)  \hfill \\
    &\;\;\;\;\;35\left( a \right)\;{\text{or}}\;35\left( b \right)\;{\text{or}}\;35\left( c \right).
\end{aligned}
\end{equation}
Note that problem (36) is again an FP problem, which can be converted to
\begin{equation}\label{eq37}
\begin{aligned}
    & \mathop {\max }\limits_{\mathbf{x}} {\text{   }}\mu g\left( {\mathbf{x}} \right) - u{{\mathbf{w}}^H}\left( {{\mathbf{\Sigma }}\left( {\mathbf{x}} \right) + {{\mathbf{I}}_{{N_R}}}} \right){\mathbf{w}} \hfill \\
    &\text{s.t.}\;12\left( b \right)\;{\text{and}}\;12\left( c \right) \hfill \\
    &\;\;\;\;\;35\left( a \right)\;{\text{or}}\;35\left( b \right)\;{\text{or}}\;35\left( c \right).
\end{aligned}
\end{equation}
One step further, since all of the constraints given in (35) are linear, the reformulation above can be tackled following the solving method proposed in Section III-C. Then, the formulation (37) is converted into a convex optimization problem which includes three subproblems. By solving the problems above, we can obtain optimal waveforms ${\mathbf{x}}_1^*,{\mathbf{x}}_2^*,{\mathbf{x}}_3^*$. Then, we substitute each of them in the objective function, the one resulting in maximum ${\text{SINR}}_{rad}$ will be the final solution to problem (36).
\subsection{With Target Location Uncertainty}
In this subsection, we study the scenario where target location is known imperfectly. Similar to Section IV, the target is assumed to locate within a given angular interval $\Psi = \left[\theta_0-\Delta\theta, \theta_0+\Delta\theta \right]$ and $\beta_p \in {\text{card}} \left(\Psi\right)$ denotes the $p$-th possible target angle. In order to guarantee the secrecy, we confine the received signal at every possible angle in the destructive area. Hence, the problem is given as follows
\begin{subequations}\label{eq38}
\begin{align}
  & \mathop {\max }\limits_{\mathbf{x}} {\text{  }}\mathop {{\text{min}}}\limits_{{\theta _p} \in {\text{card}}\left( \Psi  \right)} \frac{{\mu {{\left| {{{\mathbf{w}}^H}{\mathbf{U}}\left( {{\theta _p}} \right){\mathbf{x}}} \right|}^2}}}{{{{\mathbf{w}}^H}\left( {{\mathbf{\Sigma }}\left( {\mathbf{x}} \right) + {{\mathbf{I}}_{{N_R}}}} \right){\mathbf{w}}}} \hfill \\
  &\text{s.t.}\;\;\;\;\; {\left\| {\mathbf{x}} \right\|^2} \leq {P_0} \hfill \\
  &\;\;\;\;\;\;\;\;\; \left| {\operatorname{Im} \left( {{\mathbf{\tilde h}}_k^H{\mathbf{x}}} \right)} \right| \leq \left( {\operatorname{Re} \left( {{\mathbf{\tilde h}}_k^H{\mathbf{x}}} \right) - \sqrt {\sigma _{{C_k}}^2{\Gamma _k}} } \right)\tan \phi, \forall k \hfill \\
  & \left| {\operatorname{Im} \left( {{\alpha _0}{\mathbf{\tilde a}}_t^H\left( {{\beta _{p}}} \right){\mathbf{x}}} \right)} \right| \geq \nonumber \\
  &\;\;\;\;\;\;\;\;\;\;\;\;\; \left( {\operatorname{Re} \left( {{\alpha _0}{\mathbf{\tilde a}}_t^H\left( {{\beta _{p}}} \right){\mathbf{x}}} \right) - \sqrt {\sigma _T^2{\Gamma _T}} } \right)\tan \phi, \forall p,
\end{align}
\end{subequations}
which is however not convex. When we take the possible target locations into account, the approach proposed in Section V-A would be complicated and time consuming. Therefore, in order to reduce the computational complexity, we solve problem (38) following the given steps below. Firstly, it is noteworthy that (38d) holds when any one of the following inequalities is satisfied for each $p$.
\begin{subequations}\label{eq39}
\begin{align}
    &\operatorname{Im} \left( {{\alpha _0}{\mathbf{\tilde a}}_t^H\left( {{\beta _p}} \right){\mathbf{x}}} \right) \geq \nonumber\\
    &\;\;\;\;\;\;\;\;\;\;\;\;\;\;\;\;\;\;\;\;\;\;\;\;\left( {\operatorname{Re} \left( {{\alpha _0}{\mathbf{\tilde a}}_t^H\left( {{\beta _p}} \right){\mathbf{x}}} \right) - \sqrt {\sigma _T^2{\Gamma _T}} } \right)\tan \phi, \forall p \hfill \\
    & - \operatorname{Im} \left( {{\alpha _0}{\mathbf{\tilde a}}_t^H\left( {{\beta _p}} \right){\mathbf{x}}} \right) \geq \nonumber\\
    &\;\;\;\;\;\;\;\;\;\;\;\;\;\;\;\;\;\;\;\;\;\;\;\;\left( {\operatorname{Re} \left( {{\alpha _0}{\mathbf{\tilde a}}_t^H\left( {{\beta _p}} \right){\mathbf{x}}} \right) - \sqrt {\sigma _T^2{\Gamma _T}} } \right)\tan \phi, \forall p.
\end{align}
\end{subequations}
One step further, according to big-M continuous relaxation method proposed in \cite{cheng2013joint}, we introduce binary variables ${\eta _p} \in \left\{ {0,1} \right\},\forall p$ and a sufficiently large constant $\Omega > 0$, the reformulated either-or constraints in (39) can be converted to
\begin{subequations}\label{eq40}
\begin{align}
  &\left( {\operatorname{Re} \left( {{\alpha _0}{\mathbf{\tilde a}}_t^H\left( {{\beta _{p}}} \right){\mathbf{x}}} \right) - \sqrt {\sigma _T^2{\Gamma _T}} } \right)\tan \phi  - \operatorname{Im} \left( {{\alpha _0}{\mathbf{\tilde a}}_t^H\left( {{\beta _{p}}} \right){\mathbf{x}}} \right) \nonumber\\
  &\;\;\;\;\;\;\;\;\;\;\;\;\;\;\;\;\;\;\;\;\;\;\;\;\;\;\;\;\;\;\;\;\;\;\;\;\;\;\;\;\;\;\;\;\;\;\;\;\;\;\;\;\;\;\;\;\;\;\;\;\;\;\;\;\;\;- \eta_p \Omega  \leq 0, \forall p \hfill \\
  &\left( {\operatorname{Re} \left( {{\alpha _0}{\mathbf{\tilde a}}_t^H\left( {{\beta _{p}}} \right){\mathbf{x}}} \right) - \sqrt {\sigma _T^2{\Gamma _T}} } \right)\tan \phi  + \operatorname{Im} \left( {{\alpha _0}{\mathbf{\tilde a}}_t^H\left( {{\beta _{p}}} \right){\mathbf{x}}} \right) \nonumber\\
  & \;\;\;\;\;\;\;\;\;\;\;\;\;\;\;\;\;\;\;\;\;\;\;\;\;\;\;\;\;\;\;\;\;\;\;\;\;\;\;\;\;\;\;\;\;\;\;\;\;\;\;\;\;\;\;\;- \left( {1 - \eta_p } \right)\Omega  \leq 0, \forall p.
\end{align}
\end{subequations}
Note that in the either-or constraints above, (40a) is active when $\eta_p = 0$, which corresponds to (39a), and (40b) is fulfilled anyway due to the sufficiently large constant $\Omega$. Likewise, when $\eta_p = 1$, (40b) is activated. Accordingly, problem (38) can be recast as \cite{xu2020rethinking}
\begin{equation}\label{eq41}
\begin{aligned}
  &\mathop {\max }\limits_{\mathbf{x}} {\text{  }}\mathop {{\text{min}}}\limits_{{\theta _p} \in {\text{card}}\left( \Psi  \right)} \frac{{\mu {{\left| {{{\mathbf{w}}^H}{\mathbf{U}}\left( {{\theta _p}} \right){\mathbf{x}}} \right|}^2}}}{{{{\mathbf{w}}^H}\left( {{\mathbf{\Sigma }}\left( {\mathbf{x}} \right) + {{\mathbf{I}}_{{N_R}}}} \right){\mathbf{w}}}} \hfill \\
  &\text{s.t.}\;\;\;\;12\left( b \right), 12\left( c \right), 40\left( a \right)\;{\text{and}}\;40\left( b \right) \hfill\\
  &\;\;\;\;\;\;\;\;\; {\eta _p} \in \left\{ {0,1} \right\},\forall p,
\end{aligned}
\end{equation}
\begin{figure}
    \centering
    \includegraphics[width=0.7\columnwidth]{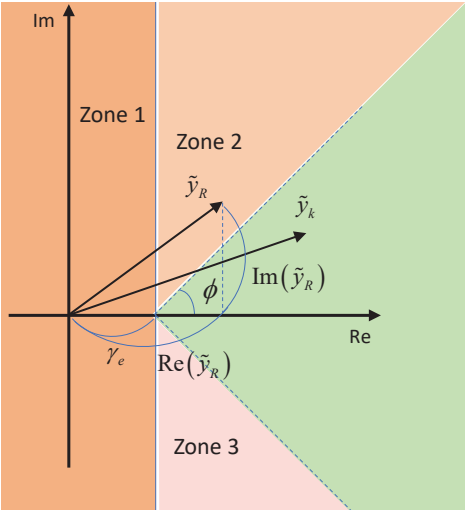}
    \captionsetup{font={footnotesize}}
    \caption{The constructive and destructive region division for QPSK.}
    \label{fig.3}
\end{figure}
We firstly reformulate the problem into the following equivalent form
\begin{equation}\label{eq42}
\begin{aligned}
  &\mathop {\min }\limits_{\mathbf{x}} {\text{  }}\mathop {{\text{max}}}\limits_{{\beta _p} \in {\text{card}}\left( \Psi  \right)} \frac{{{{\mathbf{w}}^H}\left( {{\mathbf{\Sigma }}\left( {\mathbf{x}} \right) + {{\mathbf{I}}_{{N_R}}}} \right){\mathbf{w}}}}{{\mu {{\left| {{{\mathbf{w}}^H}{\mathbf{U}}\left( {{\beta _p}} \right){\mathbf{x}}} \right|}^2}}} \hfill \\
  &\text{s.t.}\;\;\;\;12\left( b \right), 12\left( c \right), 40\left( a \right)\;{\text{and}}\;40\left( b \right) \hfill\\
    &\;\;\;\;\;\;\;\;\; {\eta _p} \in \left\{ {0,1} \right\},\forall p,
\end{aligned}
\end{equation}
\\\indent Henceforth, we will work on solving (42). Based on the formulation proposed in Section IV, we firstly give the epigraph form of problem (42), which is shown in (43).
\begin{figure*}\label{eq43}
\begin{subequations}
\begin{align}
    &\mathop {\min }\limits_{{\mathbf{x}},{\eta _p},a}  \;\;\;a  \\
    &\text{s.t.}{\text{  }}2{u_p}\sqrt {{{\mathbf{w}}^H}\left( {{\mathbf{\Sigma }}\left( {\mathbf{x}} \right) + {{\mathbf{I}}_{{N_R}}}} \right){\mathbf{w}}}  -  u_p^2\mu\operatorname{Re} \left( {{{\left( {2\left( {{{{\mathbf{x'}}}^H}{{\mathbf{U}}^H}\left( {{\theta _p}} \right){\mathbf{w}}} \right){{\mathbf{U}}^H}\left( {{\theta _0}} \right){\mathbf{w}}} \right)}^H}\left( {{\mathbf{x}} - {\mathbf{x'}}} \right)} \right) \leq a, \forall {{\theta _p} \in {\text{card}}\left( \Psi  \right)}  \\
    &\;\;\;\;\;\;12\left( b \right), 12\left( c \right), 40\left( a \right)\;{\text{and}}\;40\left( b \right) \\
    &\;\;\;\;\;\; {\eta _p} \in \left\{ {0,1} \right\},\forall p.
\end{align}
\end{subequations}
\rule[-10pt]{18.5cm}{0.05em}
\end{figure*}
It is noted that (43) is a mixed-integer optimization problem with no polynomial-time computational complexity. To reach a lower complexity, we give the equivalent form of the above problem as \cite{xu2020rethinking,ng2016power}
\begin{equation}\label{eq44}
\begin{aligned}
    &\mathop {\min }\limits_{{\mathbf{x}},{\eta _p},a}  \;\;{\text{    }}a + \omega \left( {\sum\limits_{p = 1}^{2\Delta\theta  + 1} {{\eta _p}}  - \sum\limits_{p = 1}^{2\Delta\theta  + 1} {\eta _p^2} } \right) \hfill \\
    &\text{s.t.} \;\;43\left( b \right) \hfill \\
    &\;\;\;\;\;\;12\left( b \right), 12\left( c \right), 40\left( a \right)\;{\text{and}}\;40\left( b \right) \hfill\\
    &\;\;\;\;\;\; 0 \leq {\eta _p} \leq 1,\forall p,
\end{aligned}
\end{equation}
where $\omega$ denotes a large penalty factor for penalizing the objective function for any $\eta_p$ that is not equal to 0 or 1. The problem above can solved by successive convex approximation (SCA) method firstly aiming to obtain the optimal $\eta _p$. Then, $\mathbf{x},a$ can be tackled by optimal $\eta_p$ iteratively following FP algorithm. To start with, we initially let $s\left( {{\eta _p}} \right) = \sum\limits_{p = 1}^{2\Delta \theta  + 1} {\eta _p^2}$, and the first-order Taylor expansion of $s\left( {{\eta _p}} \right)$ is given as
\begin{equation}\label{eq45}
    {\tilde s}\left( {{\eta _p}},{{\eta '}_p} \right) \approx \sum\limits_{p = 1}^{2\Delta \theta  + 1} {{{\left( {{{\eta '}_p}} \right)}^2}}  + 2\sum\limits_{p = 1}^{2\Delta \theta  + 1} {{{\eta '}_p}\left( {{\eta _p} - {{\eta '}_p}} \right)}.
\end{equation}
Herewith, problem (44) is solvable by adopting SCA algorithm so as to generate the optimal $\eta_p$. Eventually, the reformulation is given in (47), where $n$ is the iteration index of $\eta_p$. To tackle this problem, $\eta_p$ is updated until convergence, and then the optimal waveform $\mathbf{x}$ can be obtained by updating $u_p, \forall p$ iteratively by
\begin{equation}\label{eq46}
    u_p^{m + 1} = \frac{{\sqrt {{{\mathbf{w}}^H}\left( {{\mathbf{\Sigma }}\left( {{{\mathbf{x}}^m}} \right) + {{\mathbf{I}}_{{N_R}}}} \right){\mathbf{w}}} }}{{\mu {{\left| {{{\mathbf{w}}^H}{\mathbf{U}}\left( {{\theta _p}} \right){{\mathbf{x}}^m}} \right|}^2}}}.
\end{equation}
Let us denote the number of iterations required for generating the optimal $\eta_p$ by $N_{n}$. Accordingly, the total complexity of can be given as $\mathcal{O} \left( 4{N_{n}}N_T^6{\Psi _0} \right)$ by reserving the highest order term \cite{nesterov1994interior}. For simplicity, the proposed method of solving problem (41) is summarized in Algorithm 4.
\begin{figure*}\label{eq47}
\begin{align}
  &\mathop {\min }\limits_{{\mathbf{x}},{\eta _p},a} \;\;{\text{    }}a + \omega \left( {\sum\limits_{p = 1}^{2\Delta \theta  + 1} {{\eta _p}}  - \tilde s\left( {{\eta _p},\eta _p^{n - 1}} \right)} \right) \nonumber \\
  &\text{s.t.} {\text{  }}2{u_p}\sqrt {{{\mathbf{w}}^H}\left( {{\mathbf{\Sigma }}\left( {\mathbf{x}} \right) + {{\mathbf{I}}_{{N_R}}}} \right){\mathbf{w}}}  -  u_p^2\mu\operatorname{Re} \left( {{{\left( {2\left( {{{{\mathbf{x}^{m-1}}}^H}{{\mathbf{U}}^H}\left( {{\theta _p}} \right){\mathbf{w}}} \right){{\mathbf{U}}^H}\left( {{\theta _0}} \right){\mathbf{w}}} \right)}^H}\left( {{\mathbf{x}} - {\mathbf{x}^{m-1}}} \right)} \right) \leq a,\forall p \nonumber \\
  &\;\;\;\;\;\;12\left( b \right), 12\left( c \right), 40\left( a \right)\;{\text{and}}\;40\left( b \right) \nonumber \\
  &\;\;\;\;\;\;0 \leq {\eta _p} \leq 1, \forall p.  \nonumber \\
\end{align}
\rule[-10pt]{18.5cm}{0.05em}
\end{figure*}

\renewcommand{\algorithmicrequire}{\textbf{Input:}}
\renewcommand{\algorithmicensure}{\textbf{Output:}}
\begin{algorithm}
\caption{The Proposed Algorithm for solving the mixed-integer optimization problem (41)}
\label{alg:4}
\begin{algorithmic}
    \REQUIRE ${P_0},{\mathbf{h}}_k,{{\sigma _{{C_k}}^2}},{\sigma _R^2},\theta_i,b_i, \theta_0,\alpha_0, \Delta\theta, \Gamma_k, \forall k, \forall i, \varepsilon > 0, \varepsilon_0 > 0 $, and the maximum iteration number $m_{max}$
    \ENSURE ${\mathbf{x}}$
    \STATE 1. Reformulate the problem by (43).
    \STATE 2. Transform the problem to epigraph form following (31).
    \STATE 3. Initialize ${\eta _p^0 }\in \left[ {0,1} \right]$, ${\mathbf{x}}^0 \in\ \mathcal{D}$ randomly, $n = 1$, $m = 1$.
    \WHILE {$m \le {m_{max}}$ and $|| \mathbf{u}^m - \mathbf{u}^{m-1} || \ge \varepsilon $ }
    \STATE 4. When $\mathbf{x}$ is fixed, solve problem (47) iteratively by updating ${\eta _p^{n}}$ until $\left| \sum\limits_{p = 1}^{2\Delta \theta  + 1} {{\eta ^{n-1}_p}\left( {{\eta _p} - {\eta ^{n-1}_p}} \right)} \right| < \varepsilon_0$.
    \STATE 5. Fix the optimal ${\eta _p^*}$, solve problem (47) to obtain the optimal waveform ${{\mathbf{x}}^m}$.
    \STATE 6. Obtain the receive beamformer ${{\mathbf{w}}^m}$ by substituting ${{\mathbf{x}}^m}$ in (13).
    \STATE 7. Update $\mathbf{u}$ by (46).
    \STATE 8. $ m = m + 1$.
    \ENDWHILE
\end{algorithmic}
\end{algorithm}
\section{Numerical Results}
In this section, we evaluate the proposed methods via Monte Carlo based simulation results given as follows. Without loss of generality, each entry of the channel vector ${\mathbf{h}}_k$ is assumed to obey standard Complex Gaussian distribution. We assume that both the DFRC BS and the radar receiver are equipped with uniform linear arrays (ULAs) with the same number of elements with half-wavelength spacing between adjacent antennas. In the following simulations, the power budget is set as $P_0 = 30 \text{dBm}$ and the Rician coefficient is given as $v_k=1$. The target is located at $\theta_0={0^ \circ }$ with a reflecting power of ${{{\left| {{\alpha _0}} \right|}^2}} = 10 {\text{dB}}$ and clutter sources are located at $\theta_1={-50^ \circ }, \theta_2={-20^ \circ }, \theta_3={20^ \circ }, \theta_4={50^ \circ }$ reflecting a power of ${{{\left| {{\alpha _1}} \right|}^2}} = {{{\left| {{\alpha _2}} \right|}^2}} = {{{\left| {{\alpha _3}} \right|}^2}} = {{{\left| {{\alpha _4}} \right|}^2}} = 20 {\text{dB}}$. The SNR threshold $\Gamma_T$ is set as $-1{\text{dB}}$ as default unless it is presented specifically.
\begin{figure}
    \centering
    \subfigure[]{
    \includegraphics[width=0.45\columnwidth]{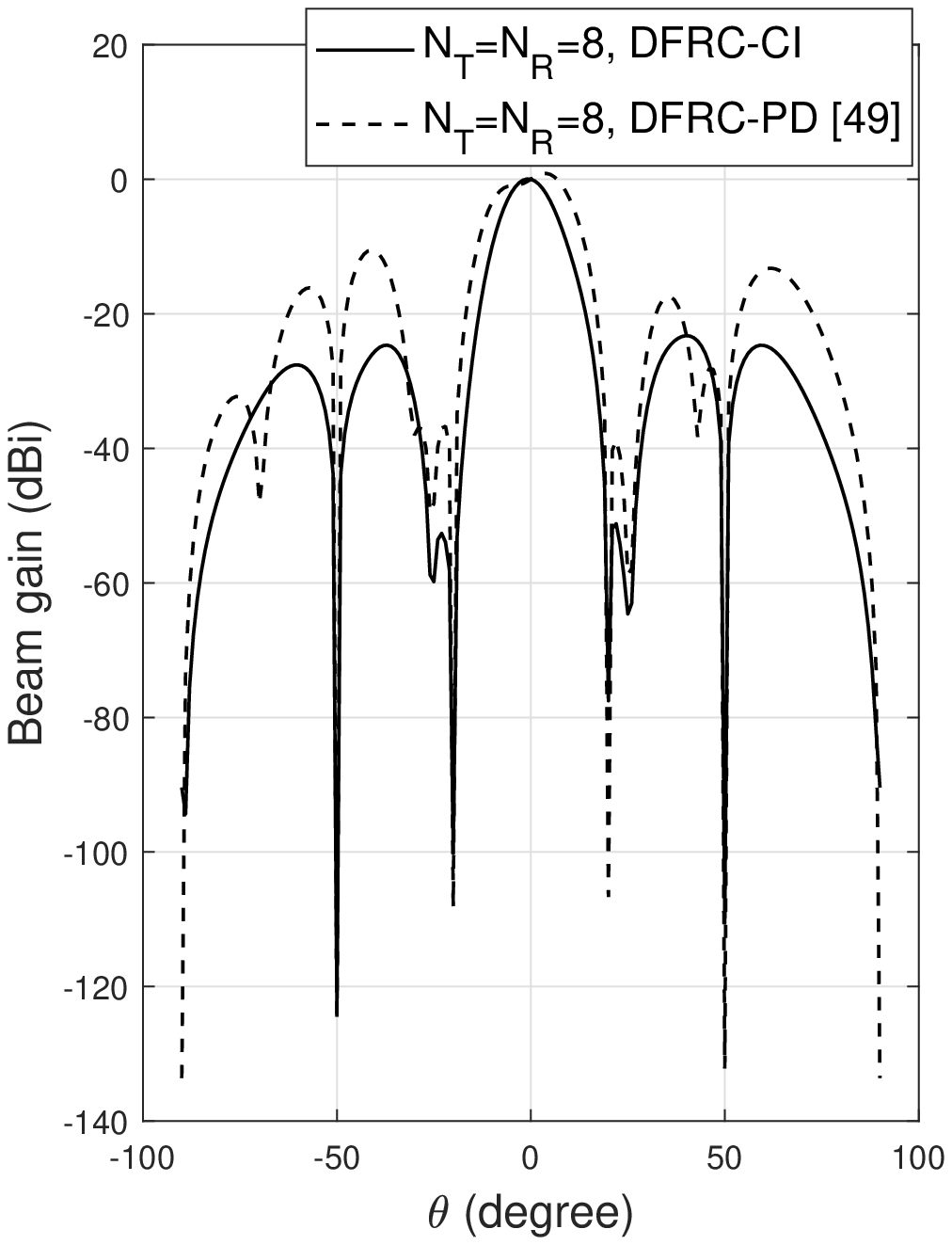}}
    \subfigure[]{
    \includegraphics[width=0.45\columnwidth]{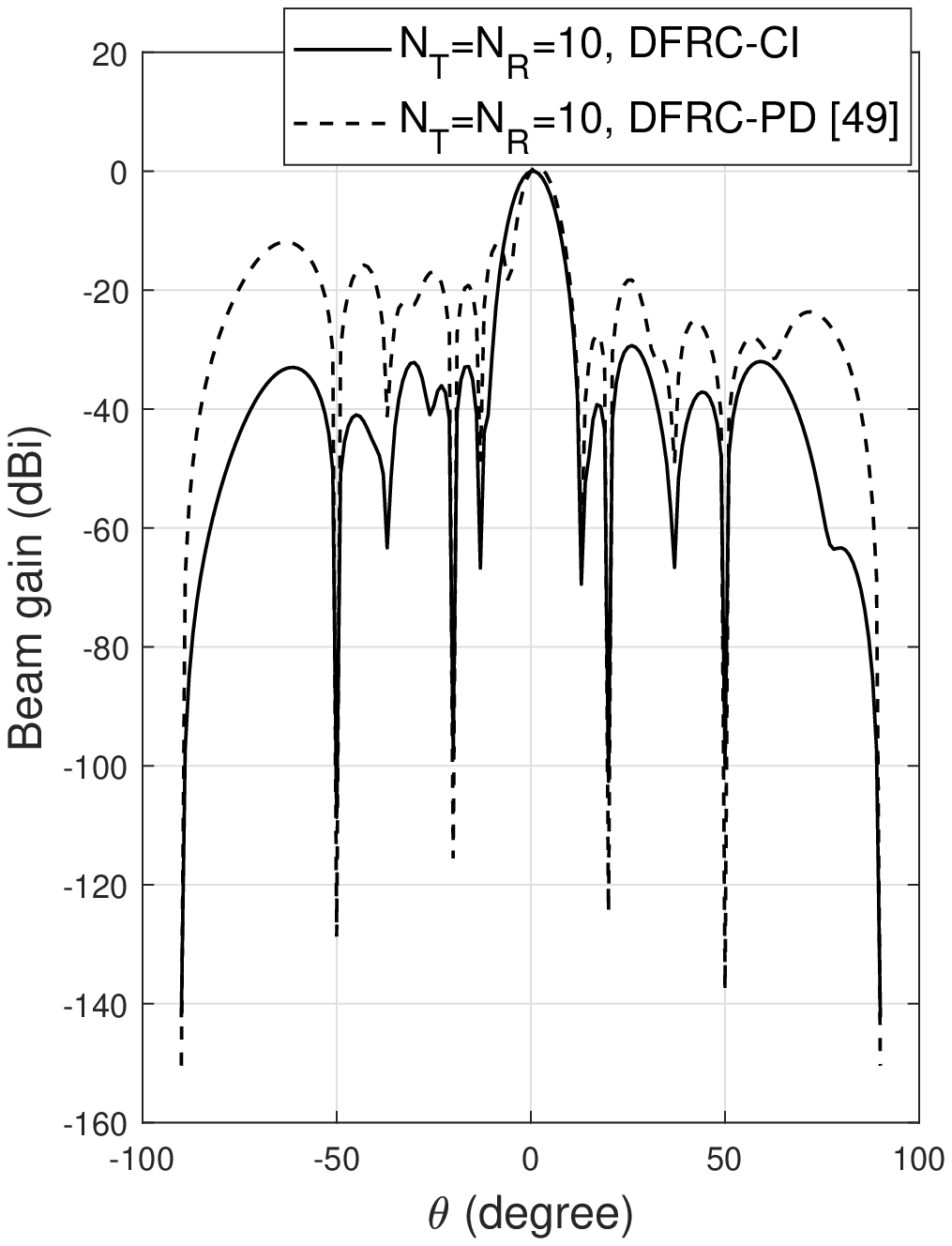}}
    \captionsetup{font={footnotesize}}
    \caption{Optimized beampatterns with different number of DFRC BS antennas, here, the beamformer design approach proposed in \cite{chen2020composite} is set as benchmarks, $K=5$.}
    \label{fig.4}
\end{figure}
\subsection{The Resultant Beampattern}
The resultant beampattern is firstly given in Fig. 4 with different number of DFRC BS antennas, where we set the DFRC precoder design proposed by Chen et al. \cite{chen2020composite} as a benchmark, namely 'DFRC-PD', and the proposed method in this paper is denoted as 'DFRC-CI' in our results. The SNR threshold $\Gamma_k, \forall k$ is fixed as 15dB. The nulls at the locations of clutter sources are clearly illustrated. It can be observed that the performance of beampattern gets better from the viewpoint of radar, and the main beam width decreases with the increasing number of BS antennas. Additionally, comparing with the beamformer design method proposed in \cite{chen2020composite}, the peak to sidelobe ratio (PSLR) of the resultant beampattern generated from our proposed waveform design method is higher, and it can be found that the null in the main beam is mitigated in our design.
\\\indent Furthermore, when the radar target location is not known to the BS perfectly, the generated beampattern is shown as Fig. 5 with different angular interval of possible target locations. It is noteworthy that the power gain of main beam reduces with the expansion of target location uncertainty interval.
\begin{figure}
    \centering
    \includegraphics[width=1\columnwidth]{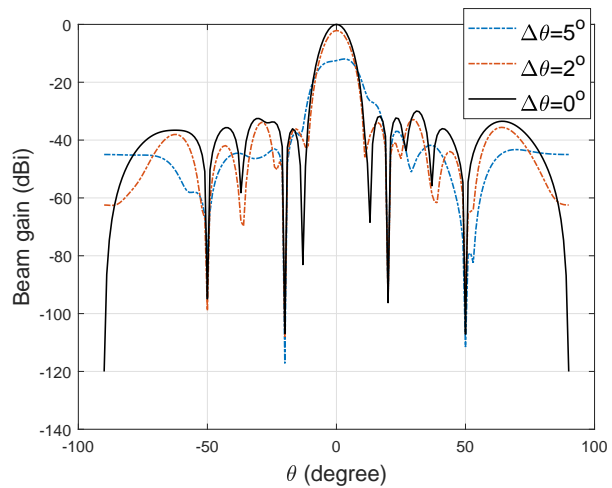}
    \captionsetup{font={footnotesize}}
    \caption{The resultant beampattern with different angular interval.$N_t=N_R=10, K=5$.}
    \label{fig.5}
\end{figure}
\begin{figure}
    \centering
    \includegraphics[width=1\columnwidth]{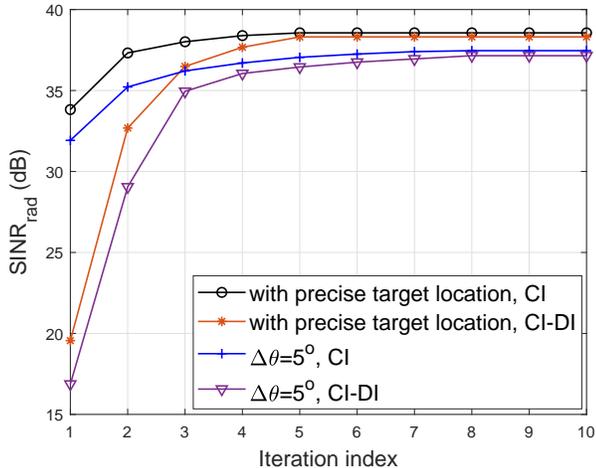}
    \captionsetup{font={footnotesize}}
    \caption{Convergence analysis.}
    \label{fig.6}
\end{figure}
\begin{figure}
    \centering
    \includegraphics[width=1\columnwidth]{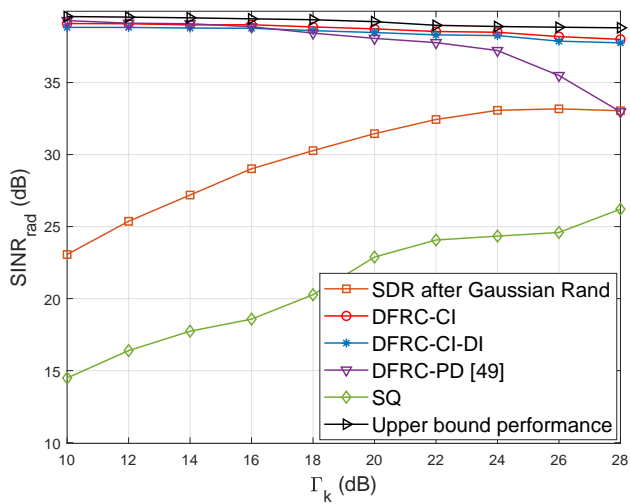}
    \captionsetup{font={footnotesize}}
    \caption{The performance of radar SINR versus CU's SNR with different solving methods, $N_T=N_R=10, K=5$.}
    \label{fig.7}
\end{figure}
\begin{figure}
    \centering
    \includegraphics[width=1\columnwidth]{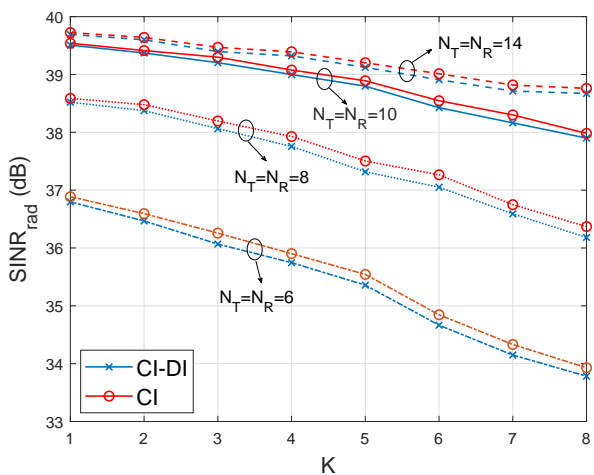}
    \captionsetup{font={footnotesize}}
    \caption{The received SINR of radar versus the number of CUs with different number of DFRC BS antennas. }
    \label{fig.8}
\end{figure}
\begin{figure}
    \centering
    \includegraphics[width=1\columnwidth]{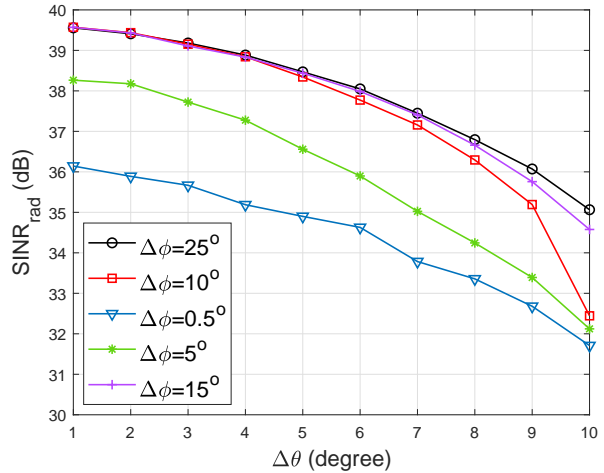}
    \captionsetup{font={footnotesize}}
    \caption{Average SINR of radar versus angular interval of target location uncertainty, $N_T=N_R=10, K=5$.}
    \label{fig.9}
\end{figure}
\subsection{Radar SINR Performance}
In this subsection, we evaluate the performance of radar receive SINR versus SNR threshold of the communication system, number of CUs, and target location uncertainty. Firstly, Fig. 6 illustrates the convergence analysis of the proposed methods. It can be found that the algorithm converges fast when the target location is precisely known to the BS. The optimal solution is generated with 5 iterations with the knowledge of precise target location, while it converges with around 9 iterations when the target location is uncertain.
\\\indent The average performance of the tradeoff between the given SNR threshold of CU and the SINR of radar is illustrated in Fig. 7, including benchmark algorithms. Specifically, with respect to the benchmarks, SQ denotes the method proposed in \cite{7450660}, SDR without Gaussian Rand denotes the upper bound of the objective function as we have given in Section III, D. To satisfy the rank-1 constraint, Gaussian randomization procedure is commonly required, and the simulation result of which is given in Fig. 7 denoted as 'SDR after Gaussian Rand'. It is found that the received SINR of radar increases with the growth of $\Gamma_k$ when we adopt SQ method and the SDR technique after Gaussian randomization procedure, while $\text{SINR}_{rad}$ decreases when we deploy the other methods. This is for the reason that the optimized system power increases with the growth of $\Gamma_k$, which is less than the given power budget $P_0$, under the circumstance when SQ method or SDR solver with Gaussian randomization procedure is deployed. That is, the SQ approach and SDR after Gaussian randomization fail to formulate an appropriate tradeoff between the radar system and the communication system. Moreover, the proposed waveform design method reaches a higher $\text{SINR}_{rad}$ comparing with the beamformer design in \cite{chen2020composite}, especially when $\Gamma_k$ is above 22dB. Furthermore, the radar receive SINR is deteriorated when the destructive interference constraints are taken into account. Fig. 8 depicts the radar SINR versus the number of CUs with different number of BS antennas, which reveals the tradeoff  between radar and communication system. It can be also noted that the receive SINR of the radar system gets lower when DI constraints are taken into account.
\\\indent In Fig. 9, we explore the effect of correlation between the target and CU LoS channels in the radar eavesdropping performance with various angular uncertainty interval $\Delta\theta$ when the angle difference between the CU and the target (i.e. '$\Delta\phi$' in Fig. 9) varies from ${0.5^ \circ }$ to ${25^ \circ }$. It indicates the tradeoff between $\text{SINR}_{rad}$ and target uncertainty. In addition, it can be found that the radar SINR is slightly impacted by the CU location when the angle difference is larger than ${15^ \circ }$.
\begin{figure*}\label{fig.10}
\centering
\subfigure[QPSK, CI]{
\includegraphics[width=6cm]{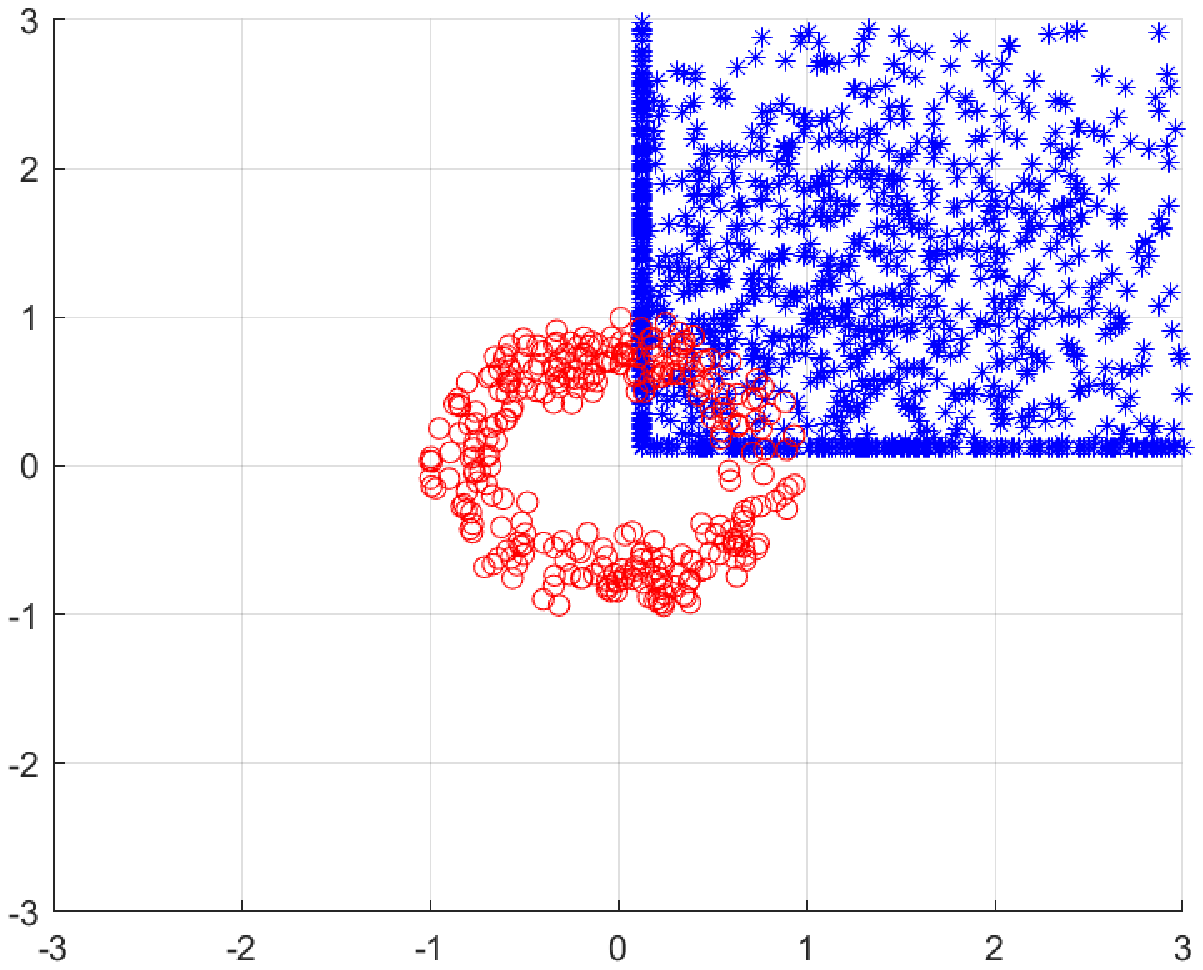}
}
\quad
\subfigure[QPSK, CI-DI]{
\includegraphics[width=6cm]{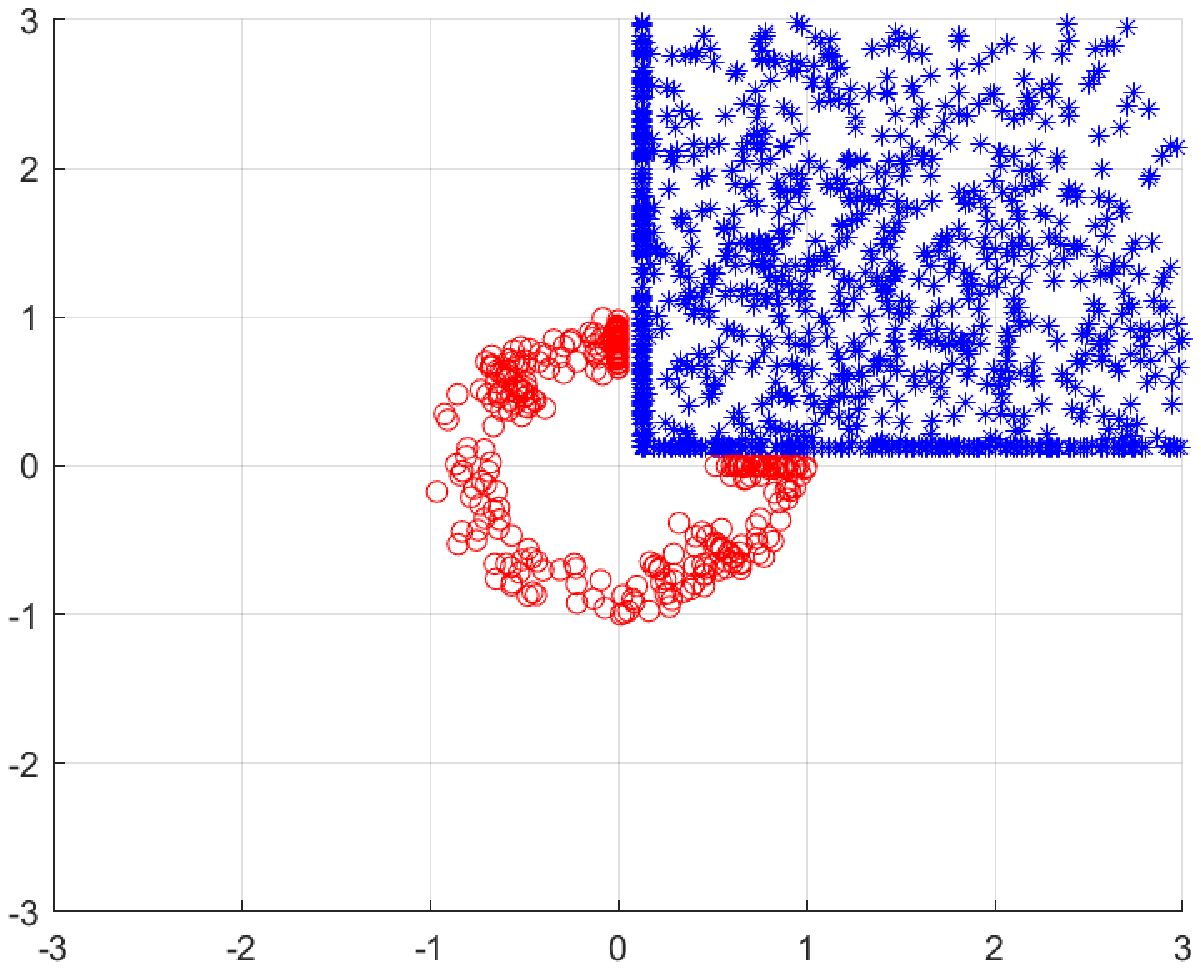}
}
\quad
\subfigure[8PSK, CI]{
\includegraphics[width=6cm]{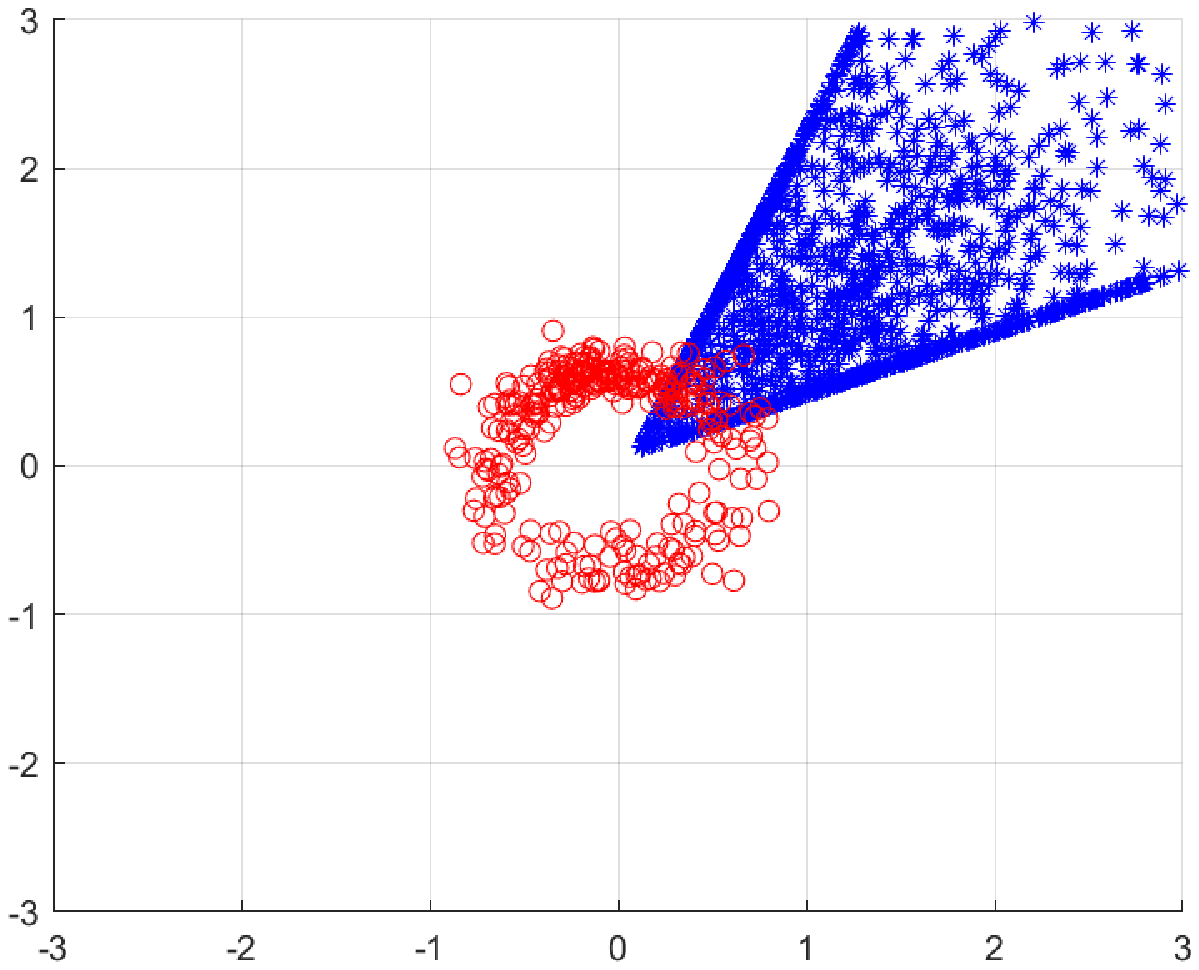}
}
\quad
\subfigure[8PSK, CI-DI]{
\includegraphics[width=6cm]{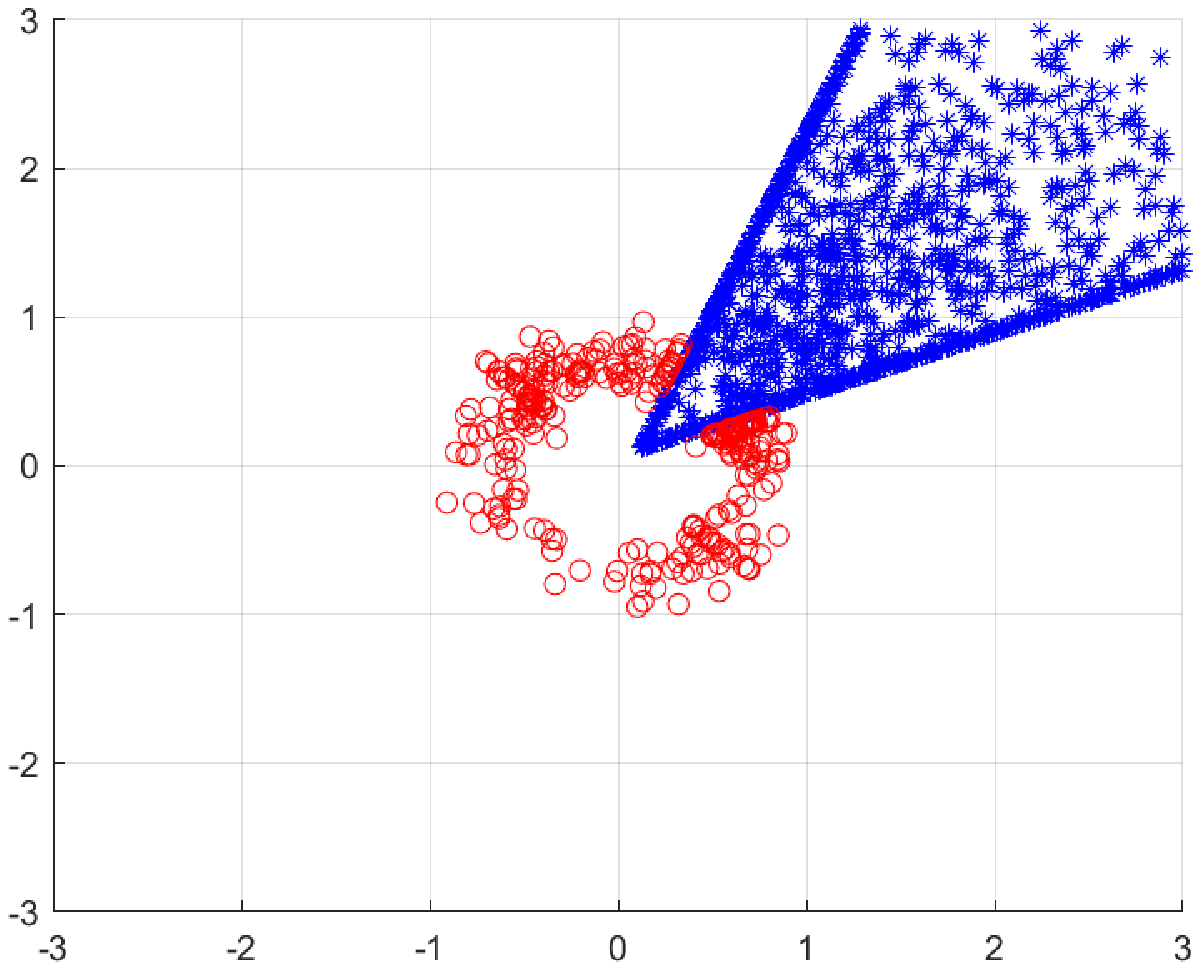}
}
\captionsetup{font={footnotesize}}
\caption{The constellation of received signals with DI constraints when the target location is known to the BS precisely, where the received signal at CUs and the target are denoted by blue dots and red dots, respectively. QPSK and 8PSK modulated signal, $N_T=N_R=10, K=5$.}
\end{figure*}
\begin{figure}
    \centering
    \includegraphics[width=1\columnwidth]{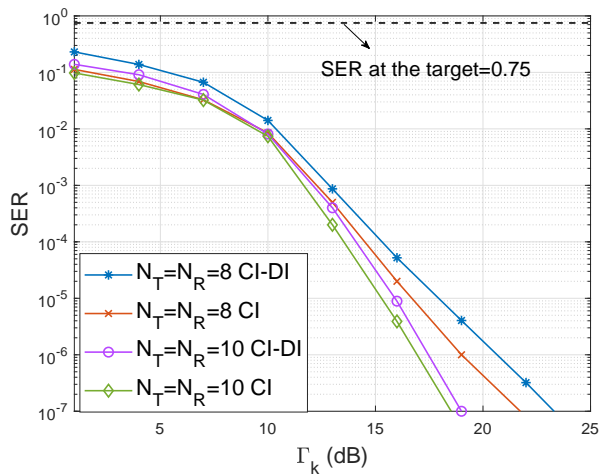}
    \captionsetup{font={footnotesize}}
    \caption{SER of CU versus SNR threshold $\Gamma_k$ with different number of antennas equipped by BS when target location is known precisely. $K=5$.}
    \label{fig.11}
\end{figure}
\begin{figure}
    \centering
    \includegraphics[width=1\columnwidth]{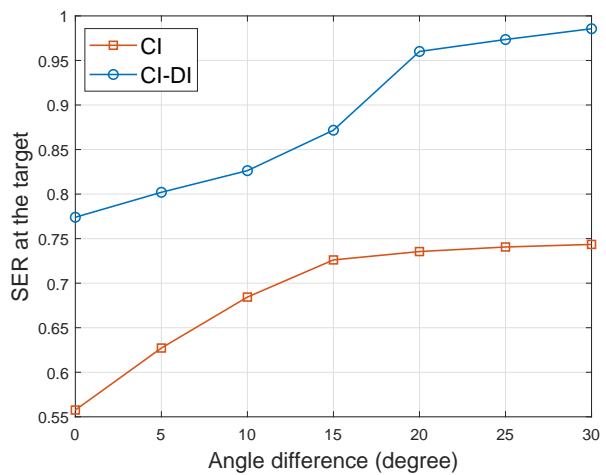}
    \captionsetup{font={footnotesize}}
    \caption{SER at the target versus the angle difference between the target and the CU with and without DI constraint when target location is known precisely. $K=5, N_T=N_R=10$.}
    \label{fig.12}
\end{figure}
\subsection{Communications Security Performance}
The distribution of received symbols at CUs (denoted by blue markers) and the target (denoted by blue markers) is shown in Fig. 10, where QPSK and 8PSK modulated symbols are taken as examples. It illustrates that the received symbols are randomized at the target when only CI is considered, while the signals received by the target are conveyed into the destructive region when deploying DI constraints. In Fig.11, the average SER of CUs versus threshold SNR $\Gamma_k$ is depicted when the BS is equipped with different number of antennas, with and without DI constraints, respectively. It is found that the SER decreases with the growth of $\Gamma_k$. Furthermore, when the received symbols at the target are constructed in the destructive region, CUs decode the received symbols with a lower probability, which means the SER performance of the CUs is deteriorated to some extend when DI constraints are taken into account.
\\\indent Furthermore, in Fig. 12, we take one CU as a reference to evaluate the SER performance of the radar target versus angle difference between the target and the CU. It is noted that target decode probability converges to 0.75 with the increasing angular difference from the CU to the target when only CI constraint is considered. For generality, the simulation result is obtained on average of target location ranging in the angular interval $\left[ { - \frac{\pi }{2},\frac{\pi }{2}} \right]$. Moreover, it can be found that the SER at the target increases obviously when the DI constraints are considered, which is close to 1 when the angle difference is getting lager. Thus, it indicates that the deployment of DI method prevents the radar target from eavesdropping communication data efficiently.
\section{Conclusion}
In this paper, we have considered the problem of secure DFRC transmission and proposed a solution based on CI. We have further extended our approach to enforce destructive interference to the target as potential eavesdropper, to further enhance security. Our numerical results have demonstrated that FP algorithms outperform the results generated from benchmark algorithms. Moreover, we observe that the DI constraints can effectively deteriorate the SER performance at the radar target, thus providing a secure solution for the unique DFRC scenarios.


\ifCLASSOPTIONcaptionsoff
  \newpage
\fi



\bibliographystyle{IEEEtran}
\bibliography{IEEEabrv,CEP_REF}

\end{document}